\documentclass[review]{elsarticle}

\usepackage{lineno,hyperref}
\modulolinenumbers[5]
\usepackage{multirow}

\usepackage[utf8x]{inputenc}
\usepackage{amsmath, amsthm, amssymb, amsfonts} 
\usepackage{graphics}
\usepackage{graphicx}
\usepackage{fancyhdr}
\usepackage{mathrsfs}
\usepackage{gensymb}

\usepackage[T1]{fontenc} 
\usepackage{xcolor}
\usepackage{listings} 
\usepackage[toc,page]{appendix}
\usepackage{comment}

\usepackage{graphicx}
\graphicspath{ {./Immagini/} }

\theoremstyle{definition}

\usepackage{listings}
\definecolor{codegreen}{rgb}{0,0.6,0}
\definecolor{codegray}{rgb}{0.5,0.5,0.5}
\definecolor{codepurple}{rgb}{0.58,0,0.82}
\definecolor{backcolour}{rgb}{0.95,0.95,0.92}

\modulolinenumbers[5]

\setkeys{Gin}{draft=false}  
\begin{document}

\begin{frontmatter}

\title{A fast computational model  for the \\ electrophysiology of the whole human heart}


\author[GSSIaddress]{Giulio Del Corso}
\author[GSSIaddress,TWENTEaddress,ROMAaddress]{Roberto Verzicco}
\author[GSSIaddress]{Francesco Viola\corref{mycorrespondingauthor}}

\cortext[mycorrespondingauthor]{Corresponding author: \textbf{francesco.viola@gssi.it}}

\address[GSSIaddress]{GSSI (Gran Sasso Science Institute), L'Aquila, Italy}
\address[TWENTEaddress]{POF Group, University of Twente, The Netherlands}
\address[ROMAaddress]{Roma Tor Vergata, Roma, Italy}

\begin{abstract}
In this study we present a novel computational model for unprecedented simulations of the whole cardiac electrophysiology. According to the heterogeneous electrophysiologic properties of the heart, the whole cardiac geometry is decomposed into a set of coupled conductive media having different topology and electrical conductivities: (i) a network of slender bundles comprising a fast conduction atrial network, the AV--node and the ventricular bundles; (ii) the Purkinje network; and (iii) the atrial and ventricular myocardium. 
The propagation of the action potential  in these conductive media is governed by the bidomain/monodomain equations, which are discretized in space using an in--house finite volume method and coupled to three different cellular models, the Courtemanche model \cite{courtemanche1998ionic} for the atrial myocytes, the Stewart model \cite{stewart2009mathematical} for the Purkinje Network and the ten~Tusscher--Panfilov model \cite{ten2006} for the ventricular myocytes.
The developed numerical model correctly reproduces the cardiac electrophysiology of the whole human heart in healthy and pathologic conditions and it can be tailored to study and optimize resynchronization therapies or invasive surgical procedures. Importantly, the whole solver is GPU--accelerated using CUDA Fortran providing an unprecedented speedup, thus opening the way for systematic parametric studies and uncertainty quantification analyses.
\end{abstract}


\end{frontmatter}

\section{Introduction}
\begin{figure}[!htbp]
    \centering
    \includegraphics[width=.8\textwidth]{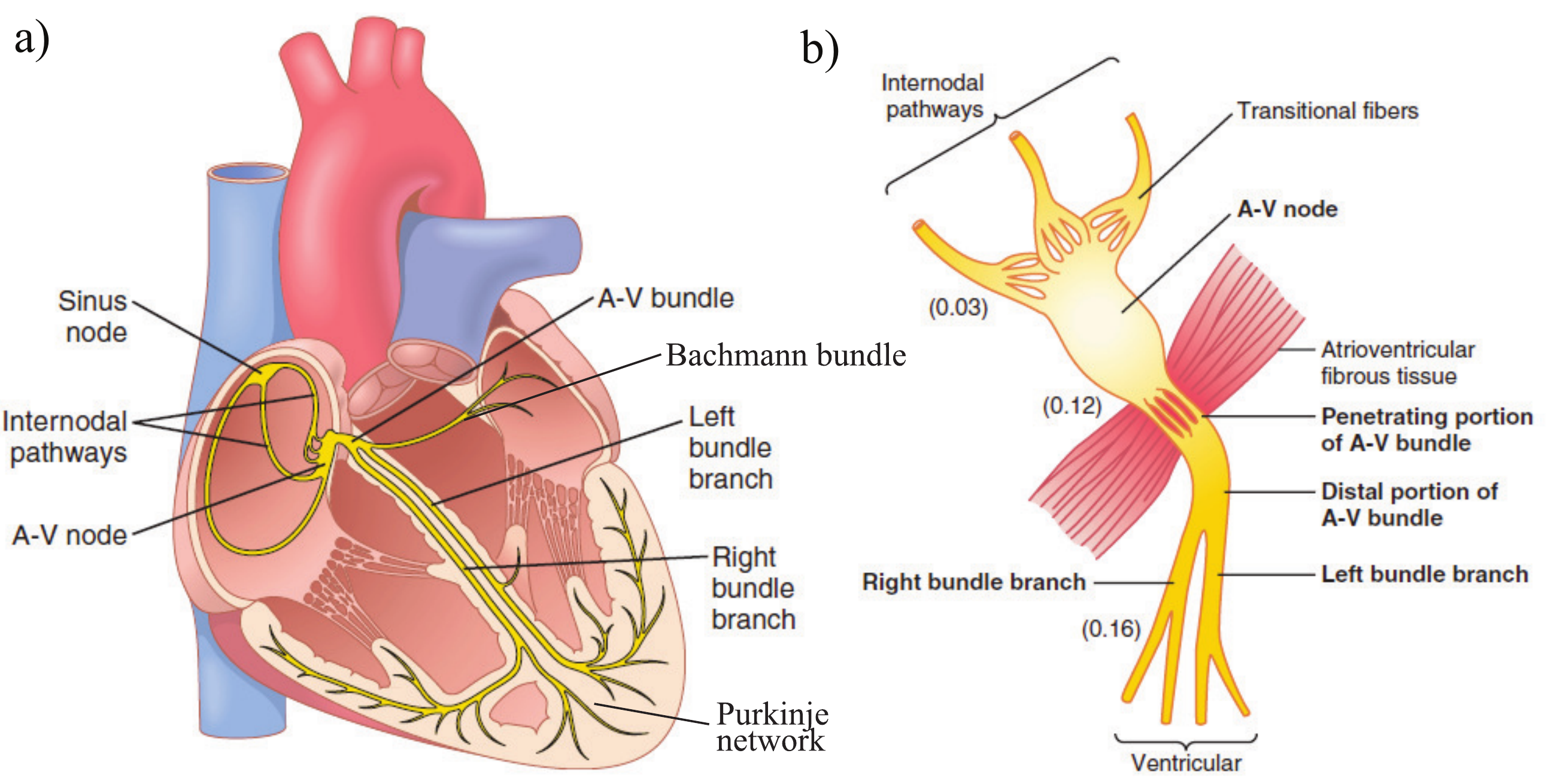}
    \caption{Sketch of the electrophysiology system of the heart \cite{hall2015guyton}. a) Fast conduction networks of bundles and Purkinje.  b) Detail of the AV bundle with the corresponding activation times (in seconds) showing the propagation delay happening in the AV--node.}
    \label{fig:GuytonRef}
\end{figure}

Owing to the development of accurate mathematical models capable of virtually replicating biological systems
and to the growing availability of computational resources to solve them, medical research is increasingly integrated with computational engineering \cite{fentonGPU2}.  
In particular, the correct modelling of the heart functioning in healthy and pathologic conditions -- such in the case of \textit{ischemic} events (reduced blood supply to a portion of the myocardium leading to dysfunction and, possibly, to the necrosis of the tissue) or of \textit{bundle branch block} (delay or blockage along the heart electrical pathway) -- entails reproducing the highly cooperative and interconnected dynamics of the heart, including its complex electrical activation. 
\\ \indent 
The latter involves many embedded conductive structures with different biological properties so as to rapidly propagate the electrical activation of atria and ventricles in order to achieve an efficient muscular contraction propelling the blood into the circulatory system.
As shown in  Figure~\ref{fig:GuytonRef}~a), the cardiac electrical depolarization, corresponding to a rise in the electrical potential across the cellular membrane owing to the transmembrane flux of ions, is initiated close to the entrance of the superior vena cava at the sinoatrial node (SA--node). Within the SA--node, some specialized pacemaker cells spontaneously produce a periodic electrical impulse, the \textit{action potential}, which propagates across the right atrium through three high speed conductivity bundles -- namely the Thorel's pathway/posterior internodal tract, the Wenckebach's middle internodal tract and the anterior internodal tract -- that wrap the right atrial chamber to assure a uniform activation.
A branch bifurcating from the latter bundle then penetrates into the internal muscle of the left atrium (Bachmann's bundle), thus initiating the depolarization also of this chamber. 
Since the propagation speed of the action potential within the fast internodal bundles is of about 1--2~m/s (significantly larger than the one observed in the atrial muscle of about 0.3--0.5~m/s \cite{hall2015guyton,harrild2000computer,james2001internodal}), after 30~ms the depolarization front reaches the atrioventricular node (AV--node) which is the electrical gate connecting the atrial with the ventricular electrophysiology system, see Figure~\ref{fig:GuytonRef}(b). 
In the AV--node, specialized cells slow down the propagation of the transmembrane potential by about 100~ms in order to allow both atria to contract before the activation wave reaches the ventricles; this avoids the simultaneous contraction of the whole organ which would produce inefficient filling/emptying of the four chambers and impaired pumping \cite{pashakhanloo2016myofiber}.
Once beyond the AV--node, the signal propagates through the His bundle, which forks into the right and left bundle branch that, in turn, progressively divide into a plethora of thin, tightly woven specialized cells named the Purkinje network, where the propagation speed of the action potential is in the range 1.5-4 m/s, corresponding to six times the propagation speed in the ventricular muscle \cite{hall2015guyton}.
This fast conduction system quickly propagates the electrical signal within the ventricular myocardium (about 30~ms to reach the terminations of the Purkinje fibers) to provide an almost simultaneous contraction of the ventricular muscle.
In addition, the Purkinje network also assures the timely activation of the papillary muscles, which stretch the \textit{chordae tendineae} so to prevent the eversion of the mitral and tricuspid valve leaflets by pulling down their free margins during early systole \cite{karas1970mechanism}.
Although the precise morphology and orientation of the Purkinje network can not be measured in--vivo, a significant variability among individuals is known to exist \cite{bordas2010integrated}, also depending on the positions of the papillary muscles which also varies among the population \cite{saha2018papillary2}. Furthermore, its smaller fibers are randomly oriented in the subendocardium with a penetration length in the myocardium of about  $0.5-100\ \mu m$ and with an average distance among them of about $0.1~$mm \cite{bergman1974atlas,tranum1991morphology}.
The Purkinje fibers are electrically isolated from the myocardial muscle, except at their endpoints called \textit{PMJs} (Purkinje Muscle Junctions), where the electrical signal can propagate from the Purkinje fibers to the ventricular myocardium with a delay ranging from to 5 to 15~ms (\textit{orthodromic} propagation) and vice--versa from the myocardium to Purkinje with a delay of 2-3~ms (\textit{antidromic} propagation)~\cite{berenfeld1998purkinje}.
\\ \indent 
Both the fast bundles and the Purkinje networks electrically activate the muscular myocardium in terms of action potential, which then propagates in the thick muscular myocardium at a lower speed that depends on the local fiber orientation. 
The myocardium is, indeed, an orthotropic medium \cite{holzapfel2009constitutive} made of oriented myocytes that enable a faster transmission of electrical impulses in the fiber direction than in the orthogonal one
and this tissue heterogeneity, playing a role in the atrial \cite{seemann2006heterogeneous} and ventricular   \cite{nickerson2005new,campbell2009effect} depolarization, should be accounted for in cardiac numerical models. 
According to the model proposed by Buckberg et al.~\cite{buckberg2008structure} the muscular fibers have a dual-orientation, with directions ranging approximately from $+60\degree$ to $-60\degree$ 
across the ventricular wall \cite{doste2019rule} and this structure has been confirmed by accurate imaging analysis of mammalians heart \cite{trayanova2011whole}. 
An additional cause of inhomogeneity is that the ventricular myocytes have different electrical properties from the atrial ones, thus resulting in a different electrical conductivity (yielding a different propagation speed) and different ionic fluxes across the myocytes membrane, which entail a different contraction pattern of atrial and ventricular chambers.
\\ \indent  
In the last decades few mathematical models for solving the cardiac electrophysiology have been proposed.
The eikonal approach solves directly the electrical depolarization of the cardiac tissue by taking as input the propagation speed within the media \cite{pullan2002finite}, whereas 
the interconnected cable methods solve the propagation of an electrical stimulus thorough a connected network of discrete cables representing the myocardium \cite{jack1975electric,leon1991directional}.
These methodologies have a limited computational cost and have been used to model the cardiac tissue including the macroscopic effects of structural heterogeneity on impulse propagation \cite{bueno2014fractional} and to incorporate more complex conduction structures, such as cardiomyocytic  fibers orientation and the His--Purkinje activation network \cite{hurtado2016computational}.  
On the other hand, leveraging on the continuum hypothesis the cardiac tissue can be modeled as an intracellular and an extracellular overlapping conductive media separated by the cell membrane. The resulting bidomain model \cite{sepulveda1989current,tung1978bi} thus consists of the coupling between a system of reaction--diffusion partial differential equations (PDEs, governing the potential propagation in the media) and a set of ordinary differential equations (ODEs) for the cellular ionic model describing the current flow through ion channels.  
The bidomain model is the state--of--the--art mathematical model for reproducing the cardiac electrophysiology  at a continuum level \cite{sundnes2007computing,vigmond2008solvers}, it has been validated against several experiments on animals \cite{wikswo1995virtual,muzikant1998validation} and it is currently adopted to solve the action potential propagation in healthy and pathologic conditions including ischemic events and fibrillation \cite{roth2001meandering,trayanova2006defibrillation,vigmond2008solvers}.
In the case the extracellular conductivity tensor is proportional to the intracellular one, the bidomain equations can be simplified into a single governing equation for the transmembrane potential, the monodomain system, which is computationally cheaper than the bidomain counterpart as the number of degrees--of--freedom (dofs of the system of PDEs) is halved \cite{sundnes2007computing}. 
Unless complex pacing patterns or fibrillation are present, the monodomain equation can be conveniently used to approximate the bidomain solution also in the case the conductivity tensors are not proportional \cite{potse2006comparison} by setting the components of the monodomain conductivity tensor to half the harmonic mean of the corresponding extracellular and intracellular components \cite{sundnes2007computing}.
\\ \indent   
The bidomain/monodomain electrophysiology model has been widely used to study different components of the cardiac electrical network such as the atrial depolarization also including pathologic atrial fibrillation \cite{wilhelms2013benchmarking,seemann2006heterogeneous}
or to model the AV--node depolarization  \cite{inada2009one,corino2011atrioventricular}.
The depolarization in the ventricular myocardium has been investigated in a series of works \cite{nickerson2005new,baillargeon2014living,sugiura2012multi,viola2020fluid} also including the 
fast conduction Purkinje network \cite{berenfeld1998purkinje,trayanova2011whole,vigmond2007construction}, which is needed to reproduce a realistic ventricular depolarization, especially in the presence of infarction  \cite{lassila2015electrophysiology} or reentry initiation of arrhythmias \cite{deo2009arrhythmogenic,deo2010arrhythmogenesis,behradfar2014role}. 
In these works, the geometry of the Purkinje network is generally obtained by applying a \textit{growing} algorithm to a one--dimensional (1D) network of fibers, which has to be sufficiently dense in order to correctly activate the 3D myocardium \cite{ijiri2008procedural,lopez2015three,vergara2016coupled}.
\\ \indent 
Although some studies are very advanced in solving the bidomain/monodomain equations in a portion of the cardiac electrical network \cite{Pathmanathan2010,Trayanova2010,Cooper2011,loppini2018competing,clayton2020audit}, a comprehensive  computational framework, solving simultaneously the fast conduction electrophysiology networks and the four--chambers muscular myocardium, is still missing. Such a computational model for the whole cardiac electrophysiology would entail, indeed, the solution of a large dynamical system, 
thus calling for efficient code parallelization with an effective use of the computational resources.
This work aims at building an accurate computational framework for solving the whole cardiac electrophysiology accounting for: (i) the fast conductivity structures of the atria and ventricles including the internodal pathways, branch bifurcations, and the AV--node; (ii) the Purkinje network immersed in the ventricular myocardium, which activates the ventricular muscle at the PMJs; (iii) the thick atrial and ventricular myocardium with their muscular fibers orientation yielding electrical anisotropy.
These three electrical components of the system have different electrophysiology properties and are modelled using a hierarchy of interconnected geometries having different topological dimension and cellular models. 
The bidomain/monodomain equations are discretized in space using an in--house finite volume method that allows for tackling complex geometries, also deforming in time, and the whole model has been ported to CUDA to run on GPU architectures thus providing unprecedented speedups \cite{fentonGPU1,fentonGPU2}.
The resulting computational model is then applied to solve the cardiac electrophysiology in healthy and pathologic conditions with the aim of assessing the model performance and validating its results.
\\ \indent
The paper is organized as follows. 
After the introduction of the cardiac geometry used throughout the work in \S~\ref{sec:domain}, the governing equations and the GPU--accelerated numerical methods are detailed in \S~\ref{sec:goveq}.
The convergence analysis of the code and validations against benchmarks results from the literature are reported in  \S~\ref{sec:convergence}. In \S~\ref{sec:results} the electrophysiology activation of the whole human heart is studied in healthy and pathological conditions, including  bundle branch blocks and the implant of artificial cardiac pacemakers. Conclusions and further research directions including possible uncertainty quantification analyses are outlined in \S~\ref{sec:discussion}.

\section{Computational domain: splitting the electrophysiology system} \label{sec:domain}
\begin{figure}[!h]
\centering
\includegraphics[width=1\textwidth]{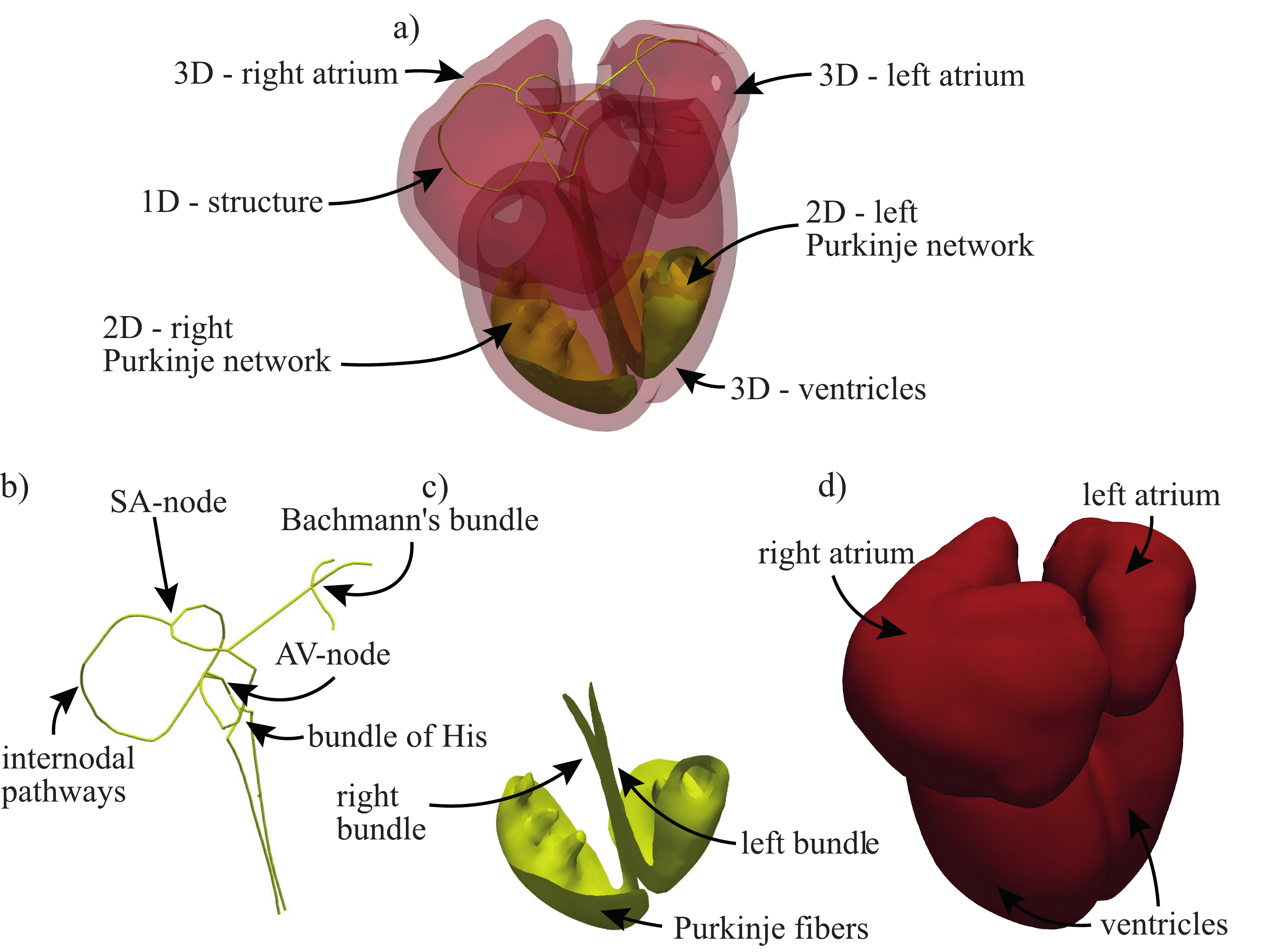}
\caption{The a) whole cardiac electrophysiology system is split in: b) 1D network of fast conduction bundles, c) 2D Purkinje network and d) 3D myocardium.}
\label{fig:subdivided}
\end{figure} 
As anticipated above, the cardiac electrophysiology system is made of a (i) fast conduction network of bundles, (ii) a Purkinje network for the ventricular activation and (iii) the massive conductive myocardium contracting as the myocytes depolarize. 
Our computational approach is based on intrinsic connections among different conductive media and pathways, 
and the complex electrophysiology system is thus split in several interconnected subdomains with different dimensional topology (see Figure~\ref{fig:subdivided}), namely a one--dimensional graph (1D) modelling the fast conduction bundles (panel~\ref{fig:subdivided} b); a two--dimensional (2D) surface approximating the dense Purkinje network (panel~\ref{fig:subdivided} c); three-dimensional (3D) media for the atrial and ventricular muscles (panel~\ref{fig:subdivided} d). 
The solution of the complete system, shown in panel~\ref{fig:subdivided}a, is thus obtained by the coupled solutions of these three distinct components which are detailed in the following.

\subsection{One dimensional fast conduction network of bundles} \label{sec:1D}
\begin{figure}[!h]
\centering
\includegraphics[width=0.7\textwidth]{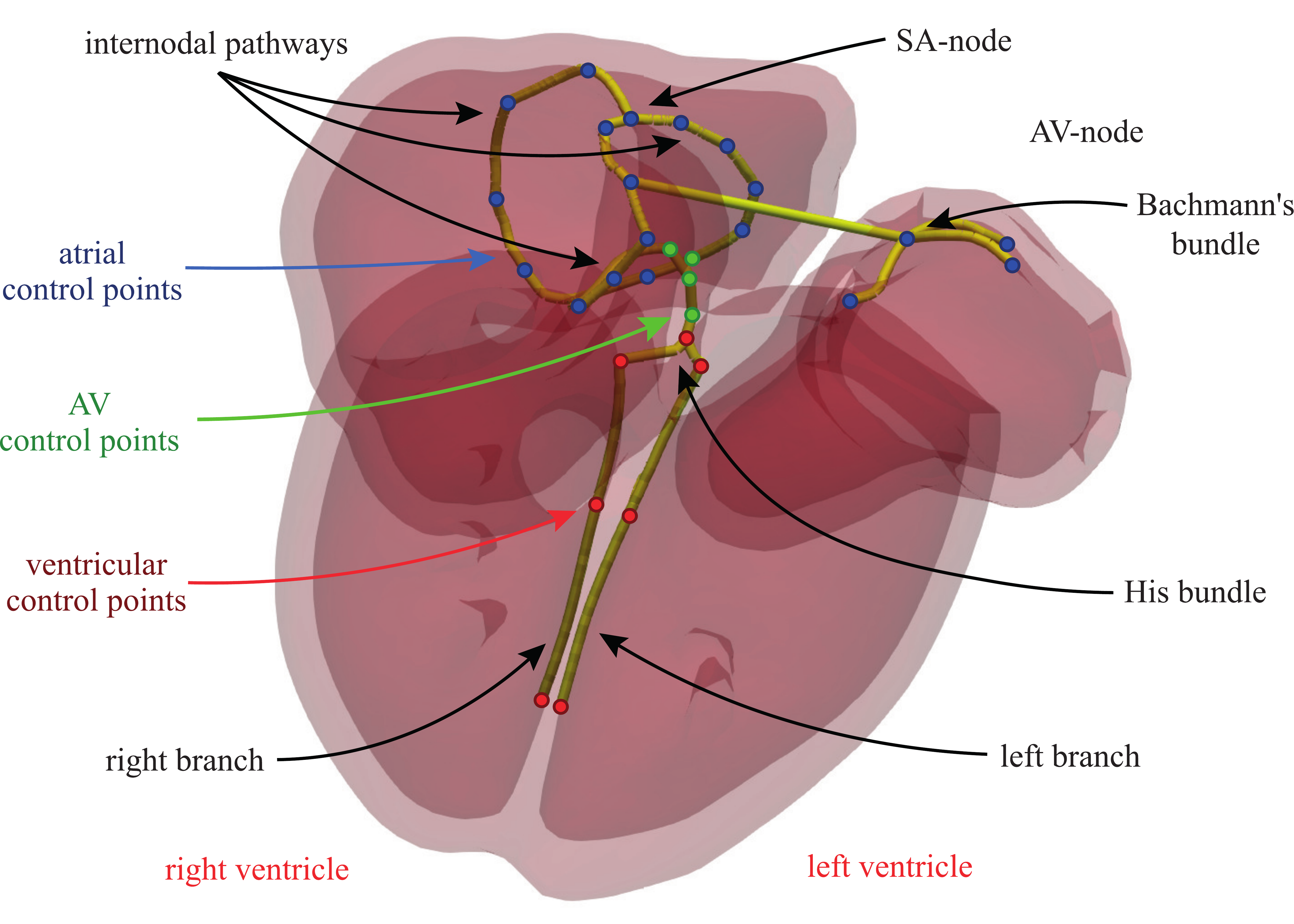}
\caption{
Fast conduction network of bundles. The circles indicate the geometrical control points of the atrial network (blue), AV--node (green) and ventricular network (red).}
\label{fig:structure1D}
\end{figure}
Owing to its slenderness, the fast conductivity structures 
conveying the electrical signal through the 3D myocardium has been modelled as a 1D fast conduction pathway with space--varying electrophysiology properties (see  Figure~\ref{fig:structure1D}).
The network originates from the SA--node and branches into the three internodal pathways reaching the AV--node with one of them (the anterior internodal pathway) further branching and connecting the right atrium to the left one through the Bachmann's bundle. 
The terminations of the internodal pathways reach the AV--node (in two locations) connecting the atrial fast conduction network with the ventricular through the bundle of His, which  then splits into two distinct branches, one immersed in the right ventricle and the other in the left one. 
\\ \indent %
In order to eventually adapt the fast conduction network to different patient geometries, the entire graph is generated through a set of control points whose coordinates can be arbitrarily set so to easily reproduce a given cardiac geometry following the adaptive procedure.
Specifically, 19 control points are distributed among the SA--node and the atrial  bundles (indicated by blue bullets in Figure~\ref{fig:structure1D}),  4 control points are used for the AV--node and its connection with the bundle of His (green bullets in Figure~\ref{fig:structure1D}) and 7 more control points are used for the ventricular bundles (red bullets Figure~\ref{fig:structure1D}).
The pathways connecting the control points are built using a piecewise linear interpolation which are then projected over the atrial and ventricular endocardium, whereas the portions of the 1D graph lying within the ventricular septum, such as the AV--node, are immersed in the 3D mesh volume.  
The 1D graph is then meshed uniformly with linear elements of a given grid size (much finer than the distance between two adjacent bullets). The whole procedure runs in few CPU--minutes, thus providing the correct positioning of a realistic 1D conduction network within the 3D mesh, with multiples bundles branching/joining the same nodes, as shown in Figure~\ref{fig:structure1D}.

\subsection{Two dimensional fast conduction Purkinje}\label{sec:purk}
\begin{figure}[!h]
\centering
\includegraphics[width=1\textwidth]{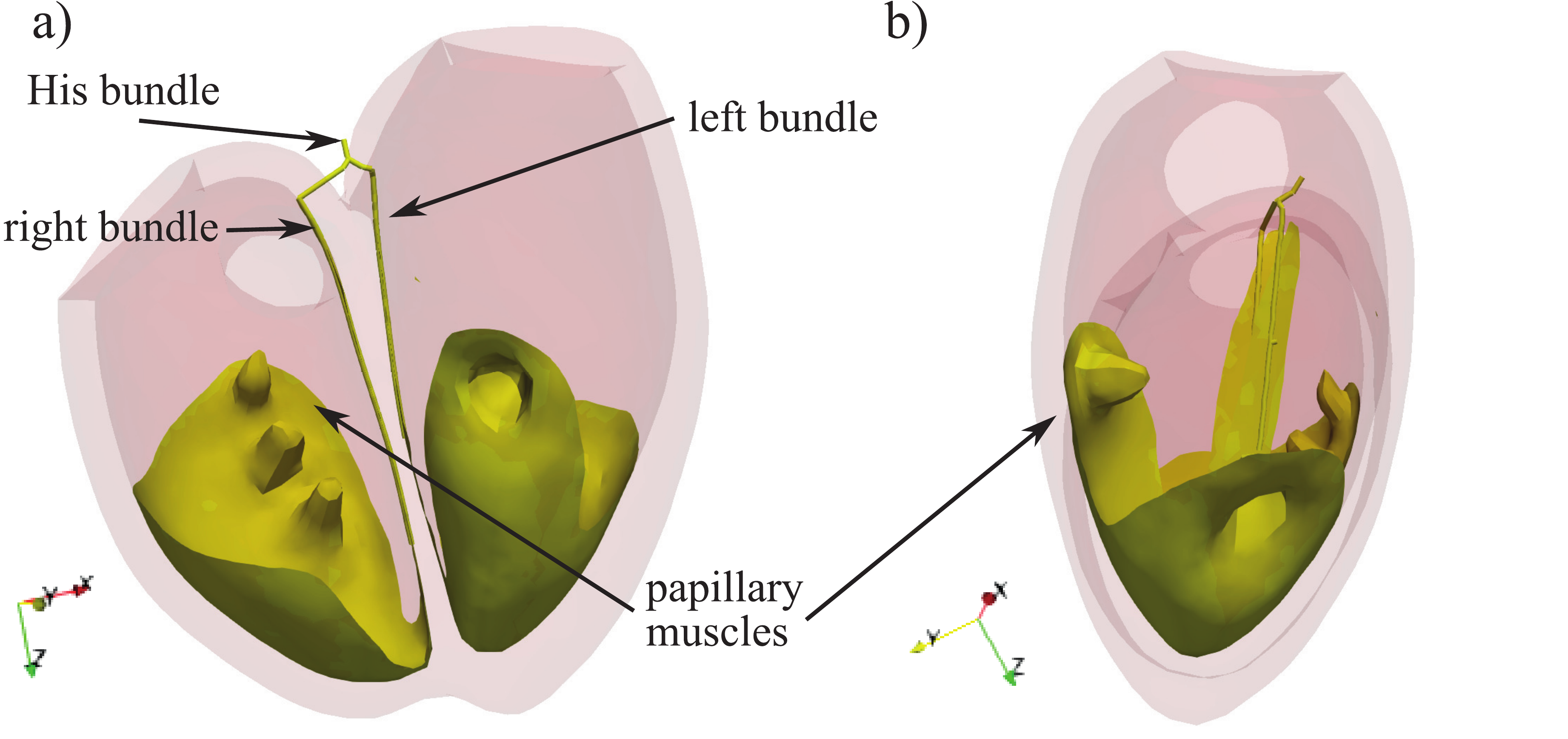}
\caption{Ventricular bundles and Purkinje network, which cover the papillary muscles.} 
\label{fig:Purkinje}
\end{figure}
The Purkinje network in humans and other mammals is distributed in a layer within the subendocardium, which is thin with respect to the myocardium thickness (of the order of 0.5--100 $\mu m$ \cite{koprla1984essential} compared to an average thickness of 7$\pm 1$ and 15.4$\pm 2.3$ mm for the right and left ventricles, respectively \cite{hutchins1978shape})
and is made of thicker fibers with a branching distance of the order of 2 mm \cite{ansari1999distribution} which bifurcate multiple times until forming a dense plethora of thinner fibers \cite{ijiri2008procedural,shimada1992purkinje}. 
This dense network of fibers is typically mimicked in computational models through the growth of a fractal structure by defining a set of generating rules and an initial topology 
(in a similar fashion to the growing models for plant branches) with the smallest branching structure in the order of  100~$\mu m$ \cite{ijiri2008procedural,bergman1974atlas,tranum1991morphology}.
As an alternative approach to the growth of a fractal 1D network, the dense fiber distribution of the Purkinje network is here merged into a continuum 2D isotropic conductive medium wrapping the endocardium.
Such approach is motivated by the uncertainty on the precise arrangement of Purkinje fibers and the great variability among individuals, which make it difficult to develop an accurate fractal rule for the network growth. 
Furthermore, a high fiber density  (more than $ 2000$ branches and 300 PMJs for the major bundles \cite{vergara2014patient} and an even smaller branching distance of 0.1--2~mm for thinner branches \cite{ansari1999distribution,ijiri2008procedural}) is required to adequately model the Purkinje and correctly activate the myocardium both in healthy  \cite{lopez2015three} and pathologic \cite{lassila2015electrophysiology} cases.
Figure~\eqref{fig:Purkinje} shows as the 2D Purkinje network develops from the His bundle and extends parallel to the left and right bundles until reaching the apex of the heart and then raises up upon two third of the ventricles height, completely covering the papillary muscles in order to timely activate their contraction at early systole.
The right and left sides of the Purkinje complex do not have a direct electrical connection since they are separated by the thickness of the interventricular septum.

\subsection{Three dimensional excitable myocardium}\label{sec:3dgeo}
The 3D myocardium is made of three excitable and conductive media, namely two  for the left and right atria and another for the ventricles (see Figure~\ref{fig:subdivided}d), which has been built using modeling software so as to reproduce high--resolution clinical images and medical atlas. This splitting of the myocardium is inspired by the cardiac electrophysiology 
as the heart septum between the atria and the ventricles (the fibrous trine plane) acts as an electrical insulator, thus decoupling the atrial and the ventricular electrophysiology. The transmembrane depolarization front, indeed, only propagates from the atria to the ventricles through the AV--node that is part of the 1D network of bundles (see \S~\ref{sec:1D}).
Similarly, the atria are electrically insulated by the atrial septum and they can thus be modelled as two disjoint electrical domains. 
On the other hand, the ventricular myocardium cannot be further subdivided into two independent meshes as, we anticipate, the ventricular endocardium is made by the same muscular fibers wrapped around the ventricles which are thus electrically connected \cite{sundnes2007computing}. 
\begin{figure}[!h]
\centering
\includegraphics[width=1\textwidth]{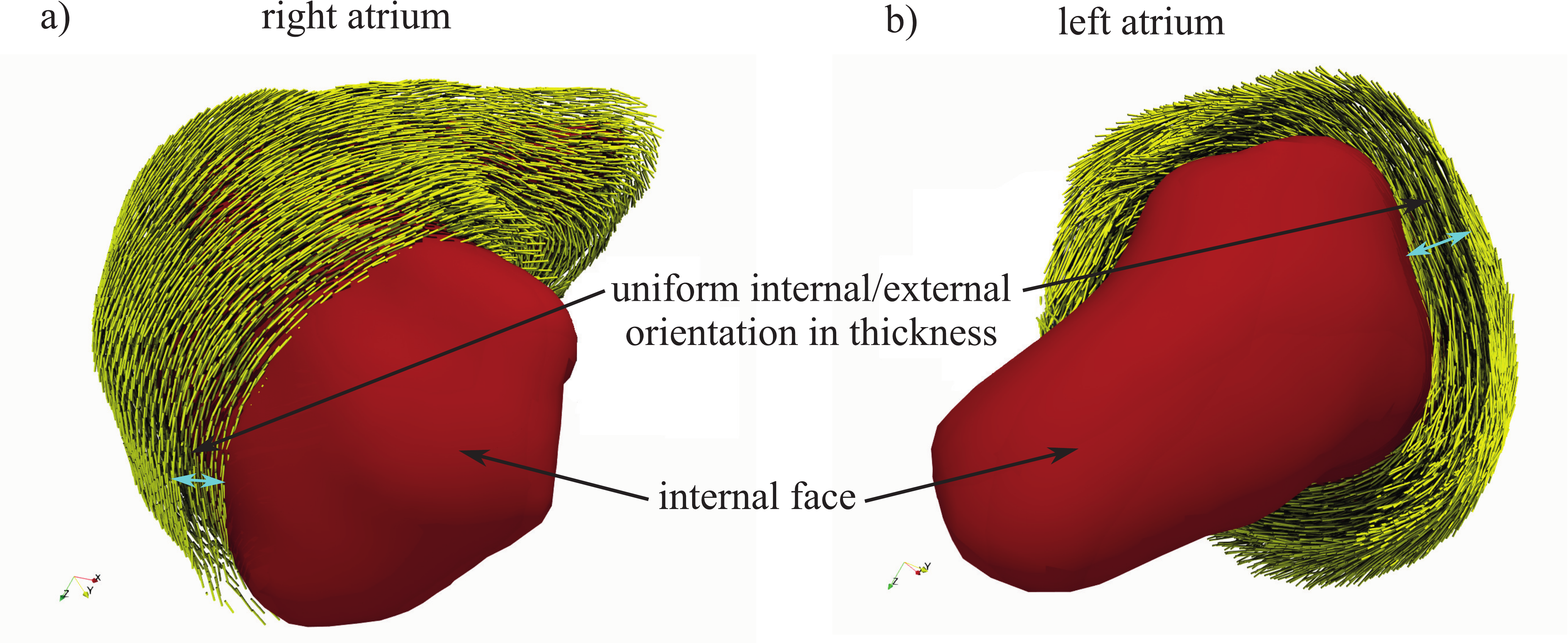}
\caption{Fibers orientation in the a) right and b) left atrium. The red surface indicate the internal endocardium.
}
\label{fig:fibersAtria}
\end{figure}
\\ \indent 
Figure~\ref{fig:fibersAtria} shows the muscular fibers orientation within the (a) right and (b) left atrial wall, with the fibers wrapping around the main atrial axes
as observed in--vivo by diffusion tensor magnetic resonance \cite{pashakhanloo2016myofiber}.  
Since the atrial fiber orientation is uniform within the myocardium thickness (of about 4~mm), this is  first defined on the atrial endocardium (red surfaces in Figure~\ref{fig:fibersAtria}) and, then replicated, at each cell across the 3D myocardium thickness.
Different or patient--specific fiber orientation in healthy and pathological conditions can be included as well in the geometrical description of the 3D media.
\begin{figure}[!h]
\centering
\includegraphics[width=0.9\textwidth]{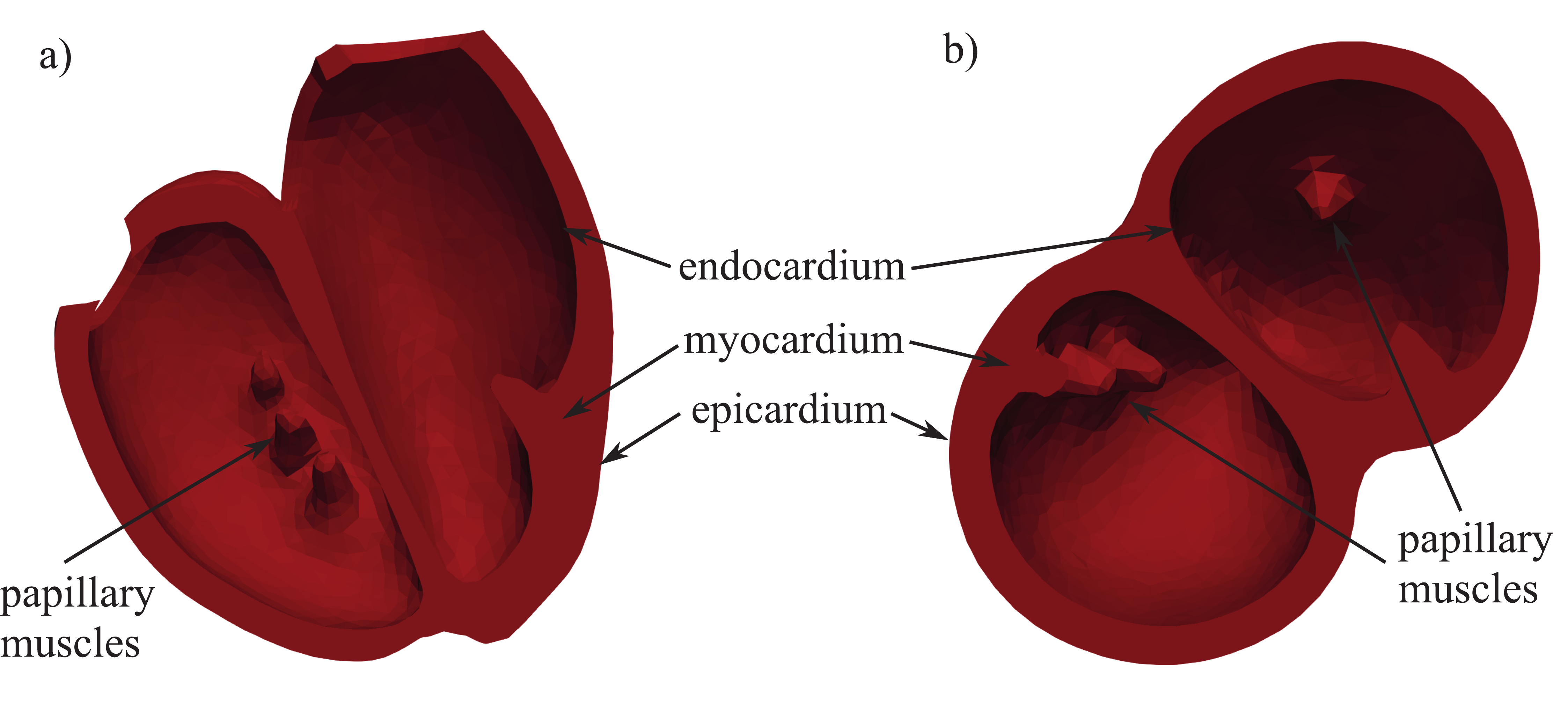}
\caption{a) Front and b) top view of the ventricular myocardium incorporating the papillary muscles.}
\label{fig:epicardium}
\end{figure}
\\ \indent 
The ventricular myocardium is modelled as a single 3D mesh for both the left and right ventricles and includes the papillary muscles, whose location corresponds to the most recurrent one observed in a population study \cite{saha2018papillary,nigri2001papillary}.
The main reason for creating a single mesh (instead of two  as in the modelling of the atria) is that the external part of the ventricular myocardium wrapping  the whole heart (often described as a \textit{scarf} \cite{buckberg2008structure}) is electrically connected and allows for a slow propagation of the depolarization front from one ventricle to the other, which is not observed in healthy cases as the two ventricles are simultaneously activated by the right and left fast conduction branches but it may occur in pathologic cases as studied in the next section. 
\begin{figure}[!h]
\centering
\includegraphics[width=0.9\textwidth]{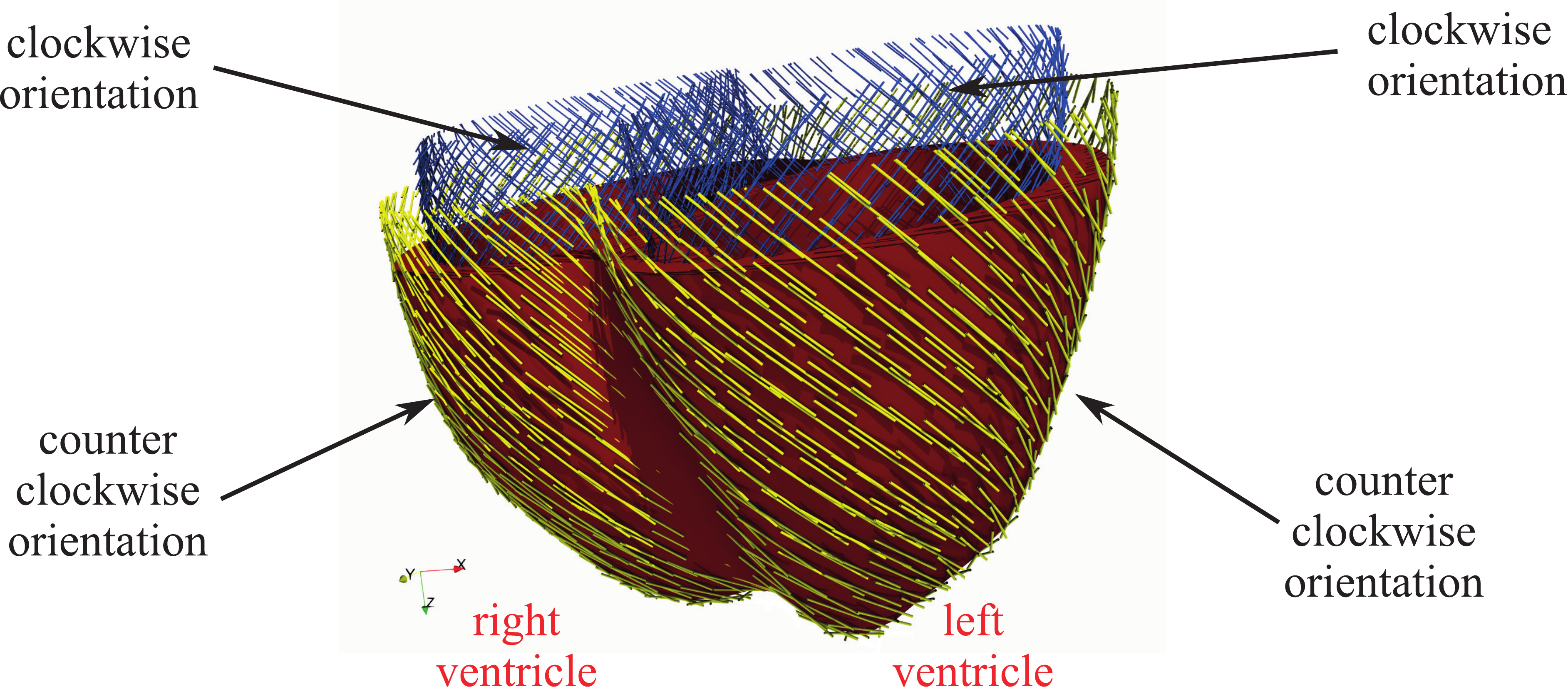}
\caption{Fibers orientation in the ventricular myocardium. The external (yellow) epicardial muscular fibers are oriented in opposite direction compared with the internal endocardial one (blue).  
}
\label{fig:3DFibers}
\end{figure} 
\\ \indent Although the orientation of the muscle fibers shows some variability among individuals,
it is known to vary across the myocardium wall from $\alpha_{epi} = 60 \degree$ at the endocardium to  $\alpha_{endo} = -60 \degree$ at the epicardium with respect to the ventricles major axis
\cite{doste2019rule,buckberg2002basic,greenbaum1981left}. 
The vector field, corresponding to the fibers orientation at each cell of the 3D mesh, is thus defined as
$\alpha = \alpha_{endo}\cdot d + \alpha_{epi}\cdot (1-d)$, where $d$ is the cell transmural distance from the endocardium normalized by the myocadium thickness (of about 8~mm on average), yielding the typical counterclockwise (clockwise) fiber orientation over the epicardium (endocadium) 
shown in Figure~\ref{fig:3DFibers}.

\section{Governing equations and numerical method} \label{sec:goveq}
\subsection{The bidomain model}
The electric wave propagating across the cardiac tissue is governed by the bidomain model that is made by the following system of two reaction--diffusion PDEs, coupled with a set of nonlinear ODEs corresponding to the cell model:
\begin{equation}\label{eq:electro}
\begin{split}
  \chi \left(   C_m \frac{\partial v}{\partial t}  +I^{ion}(v,\mathbf s) + I^s   \right)  &= \nabla \cdot (M^{int} \nabla v) + \nabla \cdot (M^{int} \nabla v^{ext}) ,  \\
    0  &=             \nabla \cdot  (M^{int} \nabla v + (M^{int}+M^{ext}) \nabla v^{ext}),            \\
    \frac{\partial \mathbf s}{\partial t} &= F(v, \mathbf s).
\end{split}
\end{equation}
Here, $v$ and $v^{ext}$ are the unknown transmembrane and extracellular potential (expressed in $mV$), whereas the surface--to--volume ratio of cells $\chi=140~\text{mm}^{-1}$  and the specific membrane capacitance $C_m=0.01 \mu \text{F mm}^{-2}$ are set as in \cite{niederer2011verification}. 
$M^{int} $ and $M^{ext}$ are the conductivity tensors of the intracellular and extracellular media that depend on the local fiber orientation with a faster propagation velocity along the fiber than in the orthogonal directions. In the case of a 3D conductive media as the myocardium these tensors have rank three and are diagonal when expressed in the fiber $(\parallel)$, sheet--fiber $(/)$ and cross--fiber $(\perp)$ directions  \cite{sundnes2007computing}, see Figure \ref{fig:3DFibers}:
\begin{equation}\label{eq:tensor}
\hat{M}^{ext} = \begin{bmatrix} m^{ext}_\parallel & 0 & 0 \\ 0  & m^{ext}_{/} & 0 \\ 0 & 0 & m^{ext}_\perp\end{bmatrix}, \hspace{1cm}
\hat{M}^{int} = \begin{bmatrix} m^{int}_\parallel & 0 & 0 \\ 0  &  m^{int}_{/} & 0 \\ 0 & 0 & m^{int}_\perp \end{bmatrix}.
\end{equation}
The conductivity tensor in the global coordinate system are thus obtained by the transformations
\begin{equation}
M^{ext} =\mathcal A \hat{M}^{ext} \mathcal A^T, \hspace{1cm}
M^{int} =\mathcal A \hat{M}^{int} \mathcal A^T,
\end{equation}
where $\mathcal A$ is the rotation matrix containing column--wise the components of fiber, sheet--fiber and cross--fiber normal unit vectors. 
On the other hand, for 2D electrical media as the Purkinje model, the transmembrane potential depolarization can only propagate in the fiber and sheet--fiber directions corresponding to the principal conductivities $m^{ext,int}_\parallel$ and $m^{ext,int}_/$. Lastly, in the case of 1D conductive media as the fast conduction network of bundles, the conduction properties are only given by the fiber conductivity $m^{ext,int}_\parallel$.
\\ \indent
The last of equations~\eqref{eq:electro} indicates the cellular model depending on the  state vector  $\mathbf s$, which couples the cellular model with the bidomain equations through the ionic current per unit cell membrane $I^{ion}$ (measured in mA/mm$^{2}$). 
Since the various components of the cardiac electrophysiology system have different cellular properties yielding different ionic fluxes and, consequently, different action potential profile, we adopt a Courtemanche cellular model \cite{courtemanche1998ionic} for the atrial myocytes (and the corresponding internodal pathways), a Stewart  model \cite{stewart2009mathematical} for the Purkinje network and  a ten~Tusscher--Panfilov model \cite{ten2006} for the  ventricular myocytes.
The ionic current, $I^s$ gives to a periodic electrical stimulus concentrated in time and space at the SA--node triggering the electrical stimulus to the ventricular myocardium, thus initiating the electrical depolarization throughout the heart:
\begin{equation} \label{eq:stimulus}
I^s = S_a ( \mathcal{H}[t] - \mathcal{H}[t-S_d] ),
\end{equation}
where $S_a = 1~mA/mm^2$ and $S_d=2.5~ms$
are the stimulus amplitude and duration, $t$ is the time within a heart beat and $\mathcal H [\cdot ]$ the Heaviside function. In a previous work, we have verified through an uncertainty quantification analysis that the values of the amplitude and duration of the stimulus do not significantly impact the subsequent depolarization of the fast conducting bundles, as far as they vary in physiological ranges \cite{del2021electrophysiology}.

\subsection{Numerical method} \label{sec:num_meth}
The set of governing equations~\eqref{eq:electro} is solved using an in--house finite volume (FV) library, which provides a suitable approach for solving the electrophysiology equation in complex geometries. As introduced above, the cardiac electrophysiology media is split in a 1D graph for the fast conduction bundles, a 2D shell for the fast conduction Purkinje and 3D media for the atrial and ventricular myocardium, which are respectively segmented with linear, triangular and tetrahedral elements.
\\ \indent
Using the divergence theorem, the bidomain equations~\eqref{eq:electro} can be rewritten in conservative form on each grid cell, $\Omega_i$,
\begin{equation}\label{eq:electro5}
\begin{split}
& \int_{\Omega_i}   \chi \left(   C_m \frac{\partial v}{\partial t}  +I^{ion} + I^s   \right) \mathrm{d}{\Omega} = \int_{\partial \Omega_i}  (M^{int} \nabla v) \cdot \mathbf n  \mathrm{d}{\gamma} + \int_{\partial \Omega_i}  (M^{int} \nabla v^{ext}) \cdot \mathbf n \mathrm{d}{\gamma},  \\
   & 0  =       \int_{\partial \Omega_i}   [M^{int} \nabla v] \cdot \mathbf n \mathrm{d}{\gamma}  + \int_{\partial \Omega_i}  [(M^{int}+M^{ext}) \nabla v^{ext})]\cdot \mathbf n \mathrm{d}{\gamma} ,            \\
\end{split}
\end{equation}
where $\mathbf n$ is the normal unit vector of the cell boundary, $\partial \Omega_i$. 

\begin{figure}[h!]
\centering
\includegraphics[width=.8\textwidth]{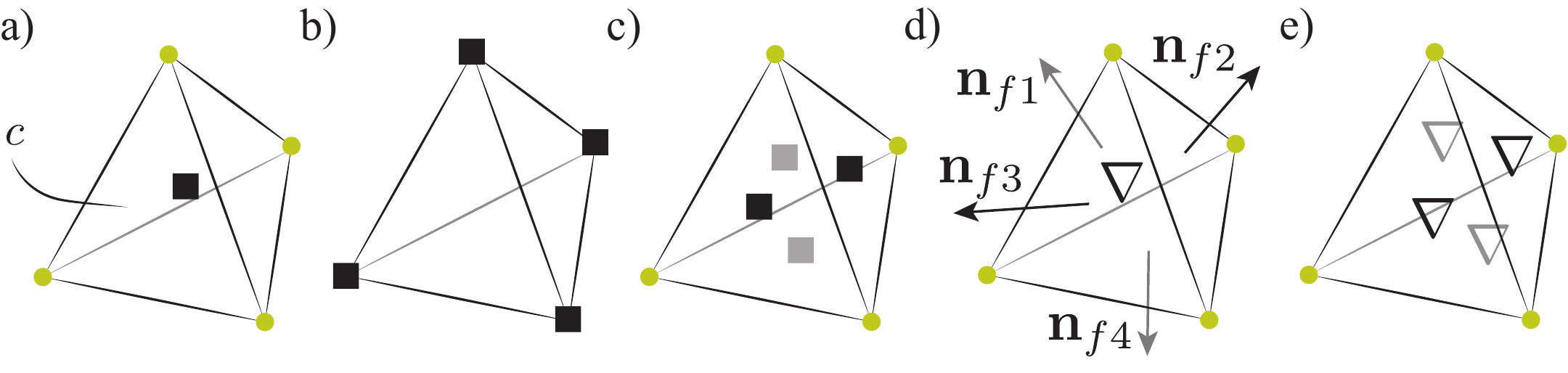}
\caption{Graphical scheme of the procedure to evaluate the gradient at the cell faces of a 3D media. a) The cell---based $v_c$ is interpolated to obtain b) the node--based $v_n$, which is then used to estimate the same quantity at c) the midpoint of the tetrahedrons faces, $v_f$. The latter is used to determine d) $\nabla v_c$ on the cell center using the Gauss--Green theorem and is successively interpolated to evaluate the e) gradient at the mesh faces $\nabla v_f$.
}
\label{fig:fv}
\end{figure}
In the case of the 3D myocardium, the domain is discretized through a tetrahedral mesh and equation~\eqref{eq:electro5} for a cell based FV method reads
\begin{equation}\label{eq:electro2_3d}
\begin{split}
&  \chi \left(   C_m \frac{\partial v_c}{\partial t}  +I^{ion}_{c} + I^{s}_{c}   \right)  V_c= \sum_{j=1}^4 A_{fj}   [ M_{fj}^{int} (\nabla v_{fj} + \nabla v^{ext}_{fj} )] \cdot \mathbf n_{fj} ,  \\
&\sum_{j=1}^4 A_{fj}   [ M_{fj}^{int}  \nabla v_{fj} ] \cdot \mathbf n_{fj} +\sum_{j=1}^4 A_{fj}   [ (M_{fj}^{int} + M_{fj}^{ext})  \nabla v^{ext}_{fj} ] \cdot \mathbf n_{fj} =0, 
\end{split}
\end{equation}
where the subscript $c$ indicates that the quantities are evaluated at the cell center whereas the subscript $fj$ denotes the $j-th$ face of the cell $c$, see \cite{fseichap}.
In the case the external and the internal conductivity tensors are parallel $M^{ext}=\lambda M^{int}$ the bidomain model~\eqref{eq:electro2_3d} reduces to the monodomain equation~\eqref{eq:electro2_3d}:
\begin{equation}\label{eq:electro2_3dmono}
\begin{split}
&  \chi \left(   C_m \frac{\partial v_c}{\partial t}  +I_{ion,c} + I_{s,c}   \right)  V_c= \sum_{j=1}^4 A_{fj}   [ M_{fj} \nabla v_{fj} ] \cdot \mathbf n_{fj} ,  \\
\end{split}
\end{equation}
where $M= \lambda M^{int}/(1+\lambda )$. 
\\ \indent
The fluxes over the tetrahedron cell faces are evaluated as indicated in Figure~\ref{fig:fv}. Firstly, the transmembrane potential at the vertex nodes $v_n$ (see panel~\ref{fig:fv}~b) is computed by 
 using the weighted average of the potential within the cells surrounding that node, $v_{k}$, yielding 
$v_n={\sum_{k=1}^{Nc_n} v_{k}d^{-1}_{k} }/{\sum_{k=1}^{Nc_n} d^{-1}_{k} }$,
where $N_{c_n}$ is the number of cells sharing the node and $d_{k} $ is the distance between the node and the $k$--th cell center. Once the values $v_n$ are found, the values of the transmembrane potential at the faces centroids $v_f$ (see panel~\ref{fig:fv}~c) are calculated by averaging the three nodal values at the triangle vertices.
According the Gauss--Green formula (panel~\ref{fig:fv}~d), the gradient of the transmembrane potential $\nabla v_c $ is related to the flux of the same quantity through the cell faces and assuming that the transmembrane potential is uniform over each mesh face we get 
$\nabla v_c  = \frac{1}{V_c} \sum_{j=1}^4 v_{fj} S_{fj} \mathbf n_{fj}$,
where $V_c$ is the volume of the cell and $v_{fj}$, $S_{fj}$, $\mathbf n_{fj},$ are the transmembrane potential, area and the normal vector at $j$--th face.
The gradient at the mesh faces is then obtained as the weighted average of the cell gradients defined at the cells $c_1$ and $c_2$ sharing the face $f$,
$\overline{\nabla v}_f  = \alpha_{c_1}  \nabla v_{c_1} + \alpha_{c_2}\nabla v_{c_2} $
where $\alpha_{c_1}$ and $\alpha_{c_2}$ are the linear interpolation weights defined on the position of the face $f$ with respect to the centers of the two cells ($\alpha_{c_1}+ \alpha_{c_2}=1$).
The resulting face gradient~$\overline{\nabla v}_f$ is computed not only using the two transmembrane potential values defined at the two cells sharing the face, but also using the cell values of all the cells sharing the nodes of  the two cells $c_1$ and $c_2$, thus enlarging the stencil of the formula. 
The 3D face gradient~$\overline{\nabla v}_f$, can be modified in such a way to include the low--stencil directional derivative and improve the stability of the method as follows 
\begin{equation}\label{eq:gra2bis}
\nabla v_f = \overline{\nabla v}_f + \left[ \frac{v_{c_1}-v_{c_2}}{d_{c_1 c_2}} - (\overline{\nabla v}_f \cdot \mathbf e_{c_1 c_2})      \right] \mathbf e_{c_1 c_2},
\end{equation}
corresponding to the last panel in Figure~\ref{fig:fv}.
The face gradient~\eqref{eq:gra2bis} can be then directly used to compute the fluxes in the conservative equation~\eqref{eq:electro} and obtain the spatially discretized bidomain equations in the 3D myocardium.
A similar FV approach is used to discretize the bidomain/monodomain equations over 1D and 2D media (in order to model the bundles and Purkinje network, respectively) with the only exception that a vertex--based FV is used in the 1D case so to better handling multiple bundles branching from the same grid node, as happening at the internodal pathway and at the Bachmann's bundle (see Figure~\ref{fig:structure1D}). 
\\ \indent 
This FV approach thus provides an effective spatial discretization of the bidomain equations over complex geometries and is second--order accurate in space provided the grid is sufficiently regular (see the convergence analysis in section~\ref{sec:convergence}). 
Importantly, as typical in FV methods the mass matrix is diagonal, thus meaning that in the case of an explicit time scheme, the discretized unsteady bidomain equation for $v$ (as well as the monodomain one)  can be marched in time simply correcting the transmembrane potential at the previous timestep by summing an incremental vector.
Although an explicit temporal scheme needs a timestep small enough to prevent numerical instabilities, still the overall computational cost is smaller than that of an implicit scheme which requires the solution of a nonlinear system at each mesh element and any timestep owing to the nonlinearity of the cellular model. 
However, the cellular models are extremely stiff, due to the significant variables variations over short timescales 
of the spike--and--dome of the action potential and of the so--called gating variables (describing the opening and closing dynamics of ion channels) and require prohibitively small timesteps to assure numerical stability.
This difficulty can be circumvented by noting that the ODEs governing the gating variables are quasi-linear  and can be solved analytically within a timestep if the transmembrane potential $v$ is held constant, whereas an explicit method is used to integrate the remaining nonlinear ones.
This semi--analytical approach is known as the Rush--Larsen scheme \cite{rush1978practical,marsh2012secrets} and it has been successfully applied to the three cellular models adopted here: the 
Courtemanche model with 15 gating variables out of 21 state variables, the Stewart model with 13 gating variables out of 20 state variables and the ten~Tusscher--Panfilov models with 13 gating variables out of 19 state variables.
The enhanced stability properties of the method thus allow for an integration timestep more than one order of magnitude larger than the one used with a standard explicit time scheme. 
\\ \indent
\textcolor{black}{
On the other hand, owing to the first order accuracy of the Rush--Larsen solution, the non--gating variables of the cell model (typically describing the variations of intracellular ions concentrations) and the spatially discretized bidomain equations~\eqref{eq:electro2_3d} are integrated in time using
 a forward Euler method \cite{marsh2012secrets}  and at each timestep the updated transmembrane potential $v(t^{n+1})$ is thus obtained as an explicit function of $v$, $v^{ext}$, $I^{ion}$ and $I^{s}$ previously computed at time $t^n$ and, similarly, the updated state vector of the cellular model $\mathbf s^{n+1}$ is computed using $\mathbf s^{n}$. As the numerical converge analysis (see section~\ref{sec:convergence}) reveals that the error of the numerical solution is more sensitive to the spatial rather than to the temporal refinement, the Rush--Larsen method with its remarkable stability properties is thus a convenient temporal scheme for the bidomain/monodomain model, although first order accurate.
Furthermore, in the perspective of multiphysics heart simulations including the coupled structural and blood dynamics, the timestep will be limited to few $\mu s$ by the fluid--structure--interaction \cite{viola2020fluid} and a first order temporal scheme for the electrophysiology system entails a numerical precision of the solution with such a small timestep. 
 } 
In the case of bidomain model, once $v(t^{n+1})$ is solved, the external potential $v^{ext}(t^{n+1})$ , is obtained by solving the linear system given by the second equation of the system~\eqref{eq:electro2_3d} through an iterative GMRES
method with restart \cite{trefethen1997numerical} using the external potential computed at previous time, $v^{ext}(t^{n})$, as first estimate for the unknown field $v^{ext}(t^{n+1})$.
\\ \indent
The FV library has been GPU  accelerated using CUDA Fortran \cite{cudabook} which extends Fortran by allowing the programmer to define Fortran functions, called kernels, which when called are executed $N$ times in parallel by $N$ different CUDA threads, as opposed to the serial nature of the regular Fortran functions, thus greatly improving the performance. Furthermore, CUDA provides CUF kernel directories which automatically run single and nested  loops on the GPU device without neither modifying the original CPU code nor writing a dedicated GPU subroutine. 
Specifically, the electrophysiology solver results in a sequence of loops on the mesh cells and on the mesh faces, which are GPU accelerated simply wrapping the original CPU code in the CUF kernel directive. 

\subsection{Subsystems coupling}\label{sec:couplingandwaterfall}
\begin{figure}[!h]
\centering
\includegraphics[width=.95\textwidth]{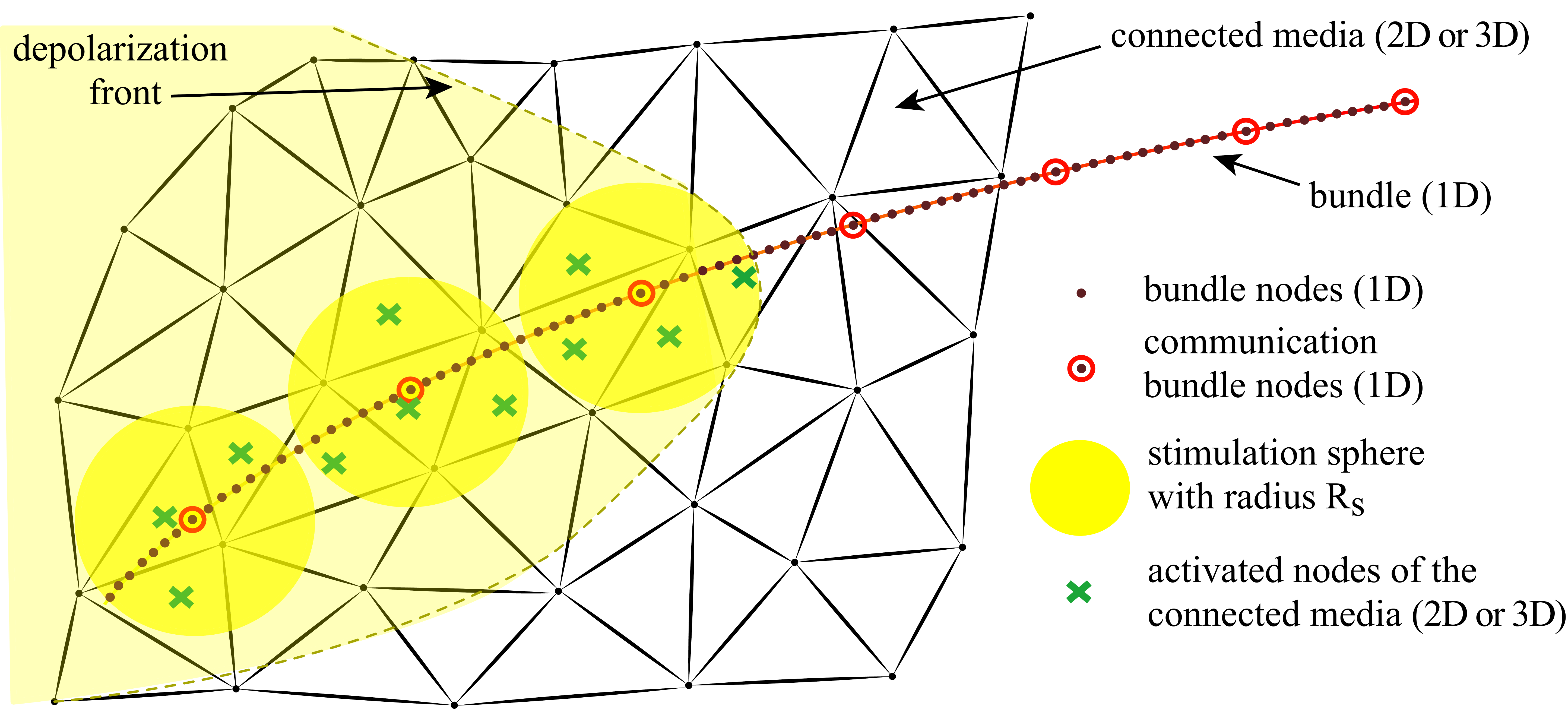}
\caption{Sketch of electrical coupling between the 1D fast conduction bundles and the surrounding 2D (or 3D) mesh.  
The wave front of the electric potential propagates across 1D mesh causing the threshold values of the communication nodes to be exceeded, thus activating the 2D (or 3D) cells within a radius $R_s$.
}
\label{fig:PropagationWave}
\end{figure}
The topological splitting of the cardiac electrophysiology network requires a coupling mechanisms to connect electrically the various subdomains. In particular, three two--way couplings are needed: (i) a first one between the 1D bundles and the 2D Purkinje networks, (ii) another between the 1D network of bundles and the 3D atrial myocardium and (iii) a last one between the 2D Purkinje and the 3D ventricular myocardium. 
\\ \indent    
As sketched in Figure~\ref{fig:PropagationWave}, the communication between the 1D mesh and the underlying 2D (or 3D) counterpart occurs through some communication nodes (CNs, indicated by red circles) which are defined in the preprocessing phase as a subset of the bundle grid nodes (black dots).
In particular, as the transmembrane potential at a CN exceeds a certain threshold (here set to 0~mV), an external localized stimulus $I^s$ (with $S_a = 1~mA/mm^2$ and $S_d = 0.5~ms$, see equation~\eqref{eq:stimulus}) is applied to the underlying 2D (or 3D) mesh cells within a distance $R_S$ from the CN, thus initiating a depolarization front in the 2D (or 3D) media.
Specifically, since the 1D domain represents the network of internodal pathways that are some millimeters thick in the atrial myocardium \cite{james2001internodal},  the communication range for the coupling between the 1D and the 3D atrial mesh is taken equal to  $R_S=1~\text{mm}$, whereas any CNs between the 1D and the 3D ventricular mesh are not present since the bundles do not directly excite the ventricular myocardium (they are isolated by fibrous sheaths) but they only transfer the propagation front to the Purkinje network  \cite{anderson2018anatomy}.
Hence, the depolarization of the Purkinje mesh is initiated by the CNs between the 1D and the 2D domains having a smaller communication range of $R_S=0.1~\text{mm}$, scaling as the local Purkinje thickness.
 Although all bundle nodes (black dots in Figure~\ref{fig:PropagationWave}) can be taken as CNs, only a subset of them is used in order to reduce the computational cost of the coupling since at any timestep the local transmembrane potential at the CNs should be monitored for eventually applying a localized electrical stimulus.
In this work, the CNs are equally distributed over the 1D network with a relative distance among them of $\varsigma  \cdot \tau \approx 1~\text{mm}$, where $\varsigma  =2$~m/s is the typical internodal pathways propagation speed and $\tau=0.5$~ms is the maximum time delay in the activation between two consecutive CNs. As a consequence, a shorter $\tau$ would correspond to a denser distribution of the CNs and vice--versa.
\begin{figure}[!ht]
\centering
\includegraphics[width=1\textwidth]{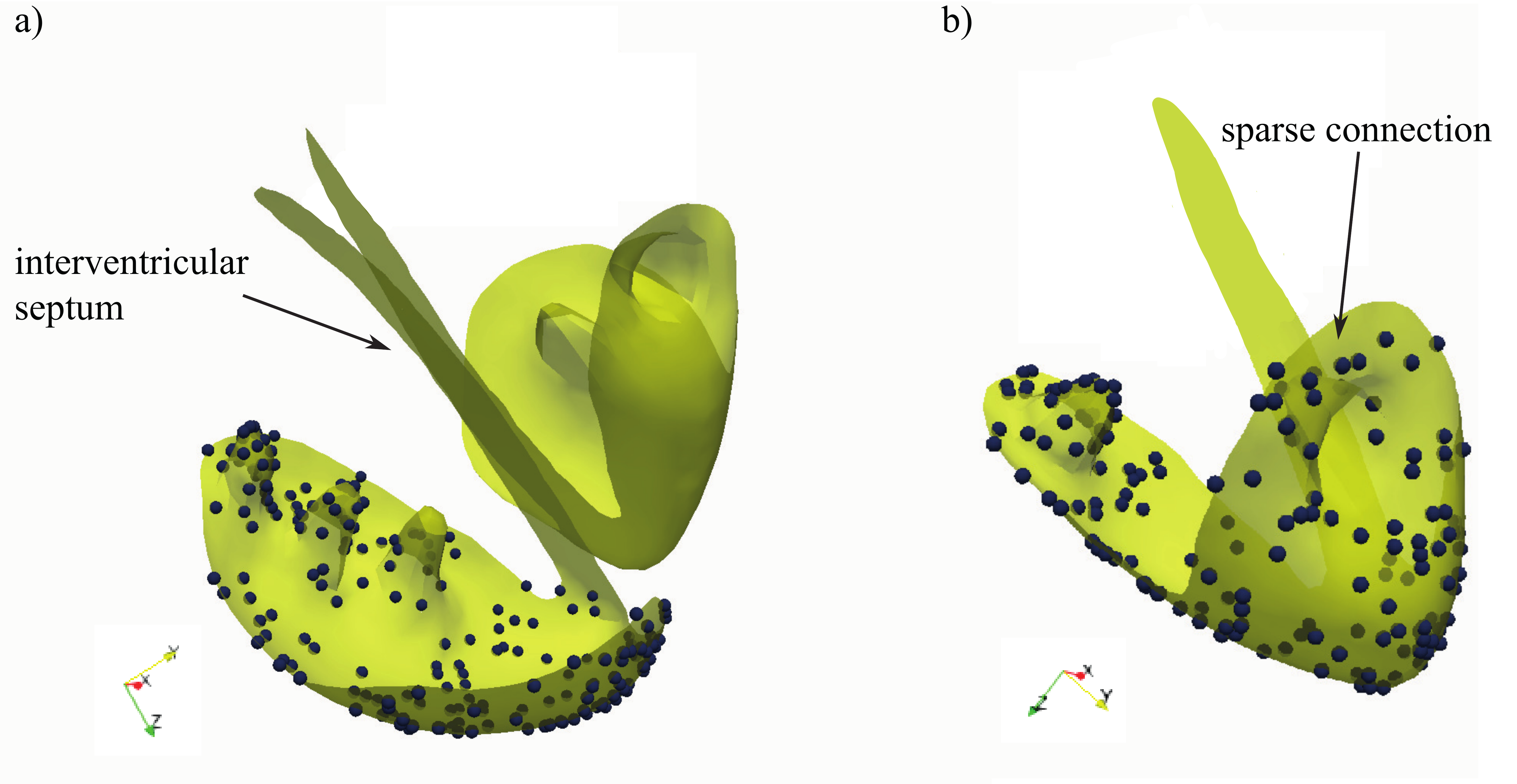}
\caption{Distribution of the communication nodes (CNs) between the 2D Purkinje and the 3D ventricular myocardium, corresponding to the Purkinje muscle junctions (PMJs).}
\label{fig:ComunicationPoints}
\end{figure}
\\ \indent Figure~\ref{fig:ComunicationPoints} show the distribution of the CNs between the 2D and the 3D ventricular media ($R_S =0.1$~mm), which allow the Purkinje network to activate the ventricular myocardium with an orthodromic delay of 5 milliseconds \cite{berenfeld1998purkinje}.
The density and the positions of these CNs is user--defined and it has been set so to reproduce the ones of the  Purkinje muscle junctions (\textit{PMJ}) \cite{vergara2014patient,ono2009morphological}. In this work 300 CNs equally distributed among the left and right ventricles \cite{liu2015image} have been considered with their distribution corresponding to the one of the PMJ with no CNs present in the interventricular septum as the Purkinje network is insulated by fibrous sheaths in that region \cite{tawara1906reizleitungssystem,anderson2018anatomy} (Figure~\ref{fig:ComunicationPoints}~a).
\\ \indent In the case of healthy cardiac electrophysiology, the electrical coupling through the CNs is \textit{one--way}, meaning that only the lower topological domain triggers an electrical stimulus on the higher one, e.g. the 1D bundle excites the 3D myocardium but not vice--versa. 
On the other hand, in some pathological cases such as nodal re-entry tachycardia \cite{narula1974sinus} or antidromic propagation (re--enter of the signal in the Purkinje network from the
myocardium) \cite{berenfeld1998purkinje}, the coupling is \textit{two--way} and the 3D myocardium can eventually excite back the 1D bundles and the 2D Purkinje, as shown in \S~\ref{sec:results}.

\section{Numerical convergence and validations} \label{sec:convergence}
\begin{figure}[!h]
\centering
\makebox[\textwidth][c]{\includegraphics[width=1.2\textwidth]{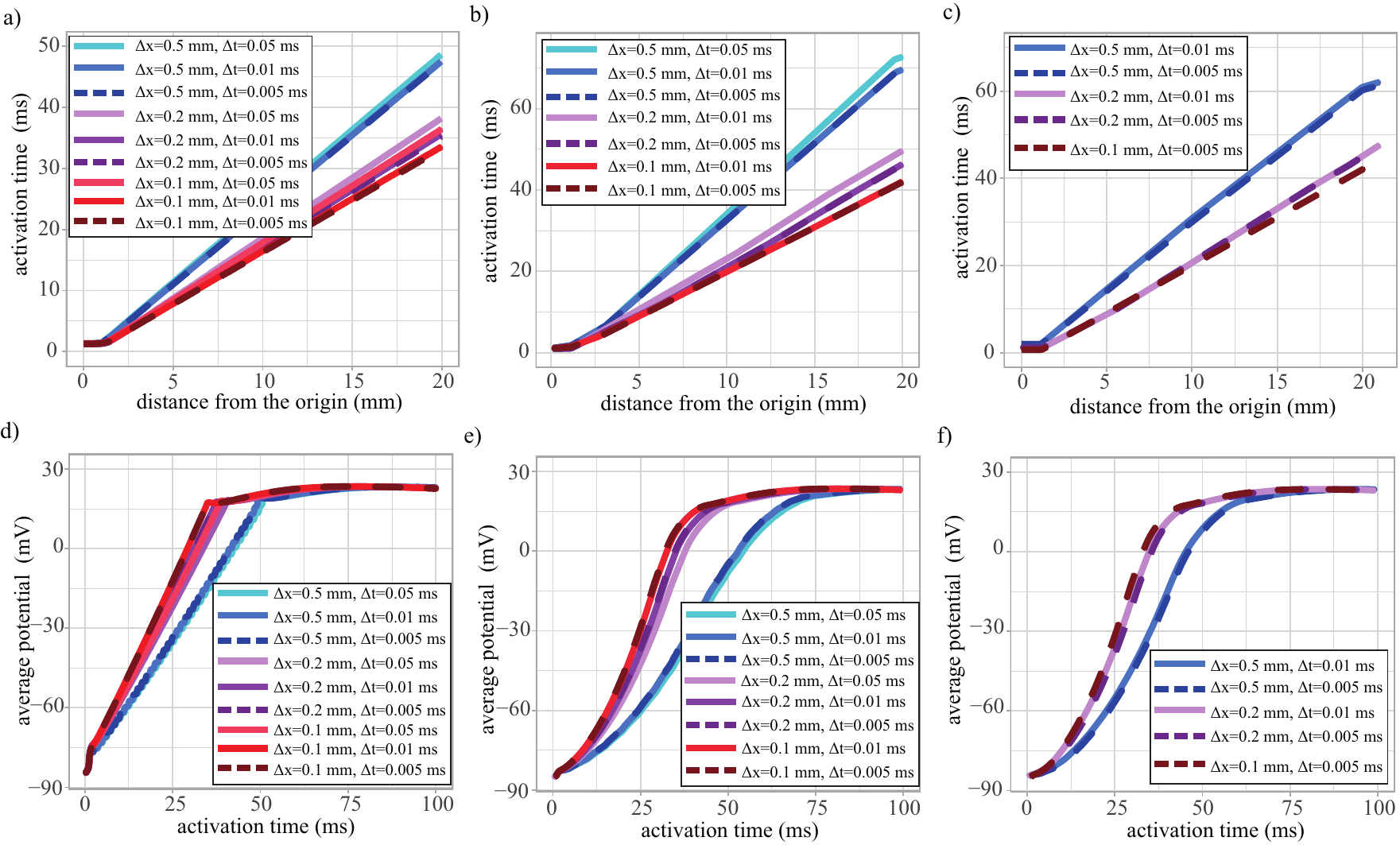}}
\caption{Activation time along the main diagonal in the a) 1D, b) 2D and c) 3D domain according to the monodomain model for various temporal and spatial resolutions ($\Delta x$ is the grid spacing the $x$ direction in 1D, $x,y$ in 2D and $x,y,z$ in 3D). The corresponding average transmembrane potential is reported in d), e) and f), respectively.   }
\label{fig:MatlabMatlab}
\end{figure}
\begin{figure}[!h]
\centering
\makebox[\textwidth][c]{\includegraphics[width=1.2\textwidth]{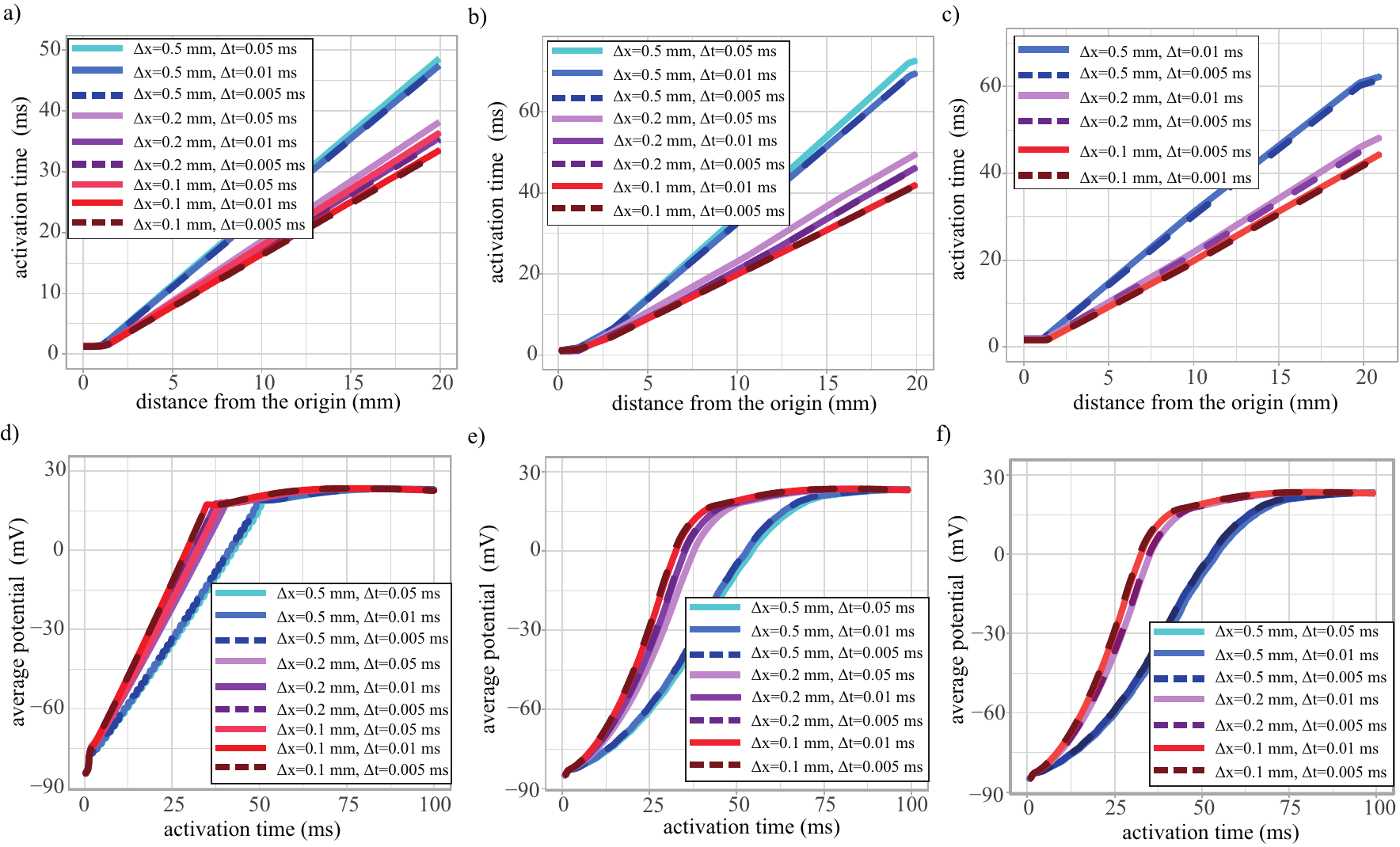}}
\caption{Same as Figure~\ref{fig:MatlabMatlab} but solving the bidomain electrophysiology model.}
\label{fig:MatlabMatlabBidomain}
\end{figure}
\begin{figure}[!t]
\centering
\makebox[\textwidth][c]{\includegraphics[width=1.4\textwidth]{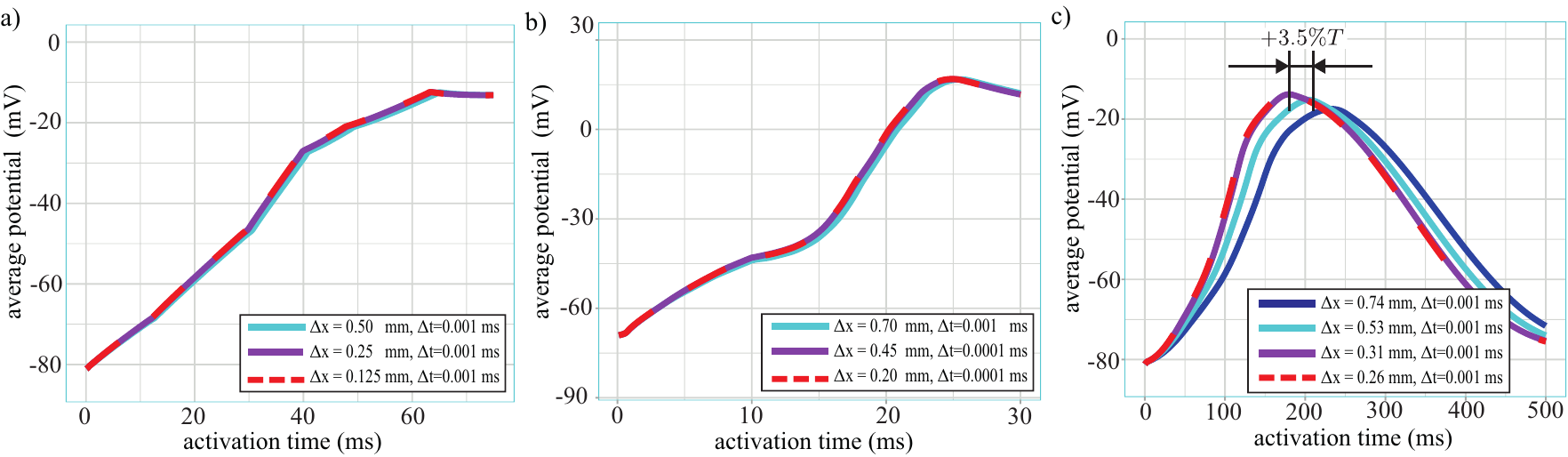}}
\caption{\textcolor{black}{Average transmembrane potential for different temporal and spatial resolutions over the a) the internodal pathways, b) the right Purkinje network and c) the left atrium. In the 3D case, a spatial resolution of $\Delta x= 0.53$~mm corresponds to a delay in the chamber activation below 3.5$\% T$, where $T$ is the heart beating period which is equal to 750~ms for a heart rate of 80~b.p.m..}} 
\label{fig:ValidationRealGeometry}
\end{figure}

The convergence of the numerical method is investigated using a procedure similar to the one reported in the benchmark paper \cite{niederer2011verification} by solving the monodomain and the bidomain equations over a 3D cartesian domain of size $20\times7\times3$ mm$^3$ coupled with the ten~Tusscher--Panfilov cellular model \cite{ten2006}.  In order to validate the 2D and 1D solvers, a similar test--case is also run on a rectangular 2D domain ($20\times 7$ mm$^2$) and on a  straight linear domain (of length 20~mm).
In all cases, the domain is discretized with a uniform spatial grid with grid size of 0.5, 0.2 and 0.1 mm in each direction ($x$ in the 1D, $x,y$ in the 2D and $x,y,z$ in the 3D), and three different timesteps have been used, namely 0.05, 0.01 and 0.005 ms. 
The muscle fibers are taken aligned with the long axis direction (20 mm in 2D and 3D) and the electrophysiology parameters, including the initial state variables of the cell model, are set as in \cite{niederer2011verification}. The initial stimulus is applied within a line/square/cube of side 1.5 mm placed in the corner closer to the origin. 
\\ \indent 
In the case of the monodomain solver, Figure~\ref{fig:MatlabMatlab} reports the activation time (defined as the instant when the transmembrane potential exceeds~0~mV)  along the diagonal of the domain departing from the corner where the stimulus is applied for the (a) 1D (b) 2D and (c) 3D domains. The corresponding transmembrane potential averaged in the domain volume $V$, $ \overline{v}(t) = \int_{V} v(\mathbf x,t) dV / V $ are  reported as a function of time in Figure~\ref{fig:MatlabMatlab}(d,e,f), showing that, as the spatial grid is refined, the propagation speed of the depolarization front increases until convergence is attained for the more refined grid spacing with a timestep equal--smaller than $\Delta t = 0.01$~ms.
The corresponding convergence curve for the bidomain model are given in Figure~\ref{fig:MatlabMatlabBidomain} for each topological dimension of the conductive media, thus showing similar results for the same grid spacing and timestep.
\begin{figure}[!h]
\centering
\includegraphics[width=1\textwidth]{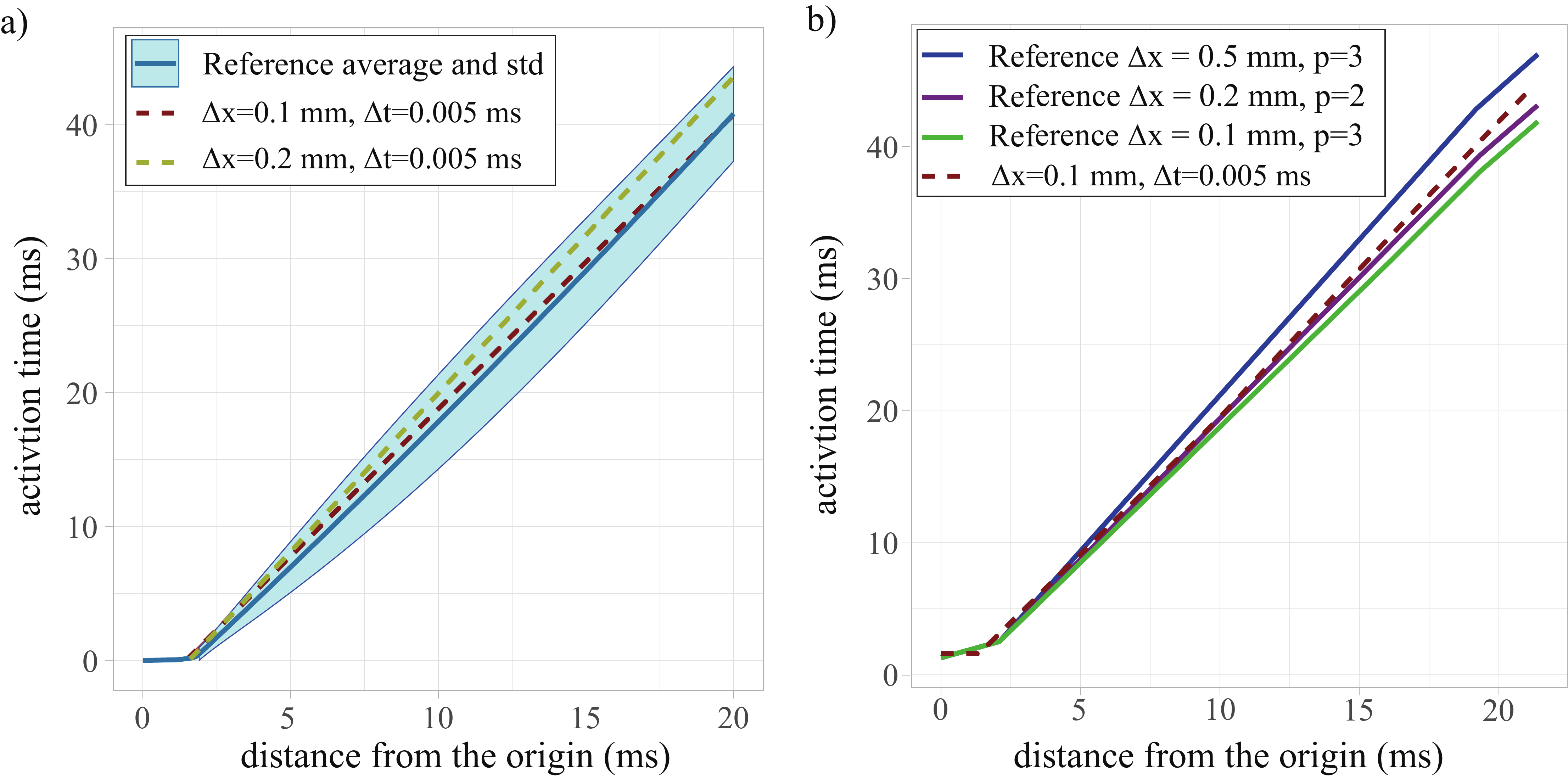}
\caption{Average transmembrane potential in the 3D domain according to a) the monodomain and b) the bidomain electrophysiology model. a) Our numerical results (yellow dashed line for $\Delta x=0.2$ mm and red dashed one for $\Delta x = 0.1$ mm) are compared against the average results (blue solid line) of the benchmark paper \cite{niederer2011simulating}, where the shaded area indicates the corresponding standard deviation. 
In panel b) our code is validated against the results reported \cite{cuccuru2015simulating}. 
} 
\label{fig:MatlabPython}
\end{figure}
\\ \indent
In the 3D case, the numerical solution of the monodomain and bidomain equations can be validated against previous results from the literature.  
The former is validated against the benchmark paper of Niederer et al.~2011 \cite{niederer2011verification} where 11 different numerical codes (either based on finite elements or finite differences) have been used and the resulting average activation time along the diagonal of the cubic domain (blue solid line) and standard deviation (blue shaded region) are reported in Figure~\ref{fig:MatlabPython}~(a).  Moreover, the solution of the bidomain equations is validated against the results of Cuccuru et al. \cite{cuccuru2015simulating}, which is reported in  Figure~\ref{fig:MatlabPython}~(b) for different spatial steps and different polynomial degrees, see \cite{cuccuru2015simulating} for further details on their numerical method.
Both the monodomain and the bidomain results obtained with our numerical solver fit well those reported in the literature.
\begin{figure}[!h]
\centering
\makebox[\textwidth][c]{\includegraphics[width=1\textwidth]{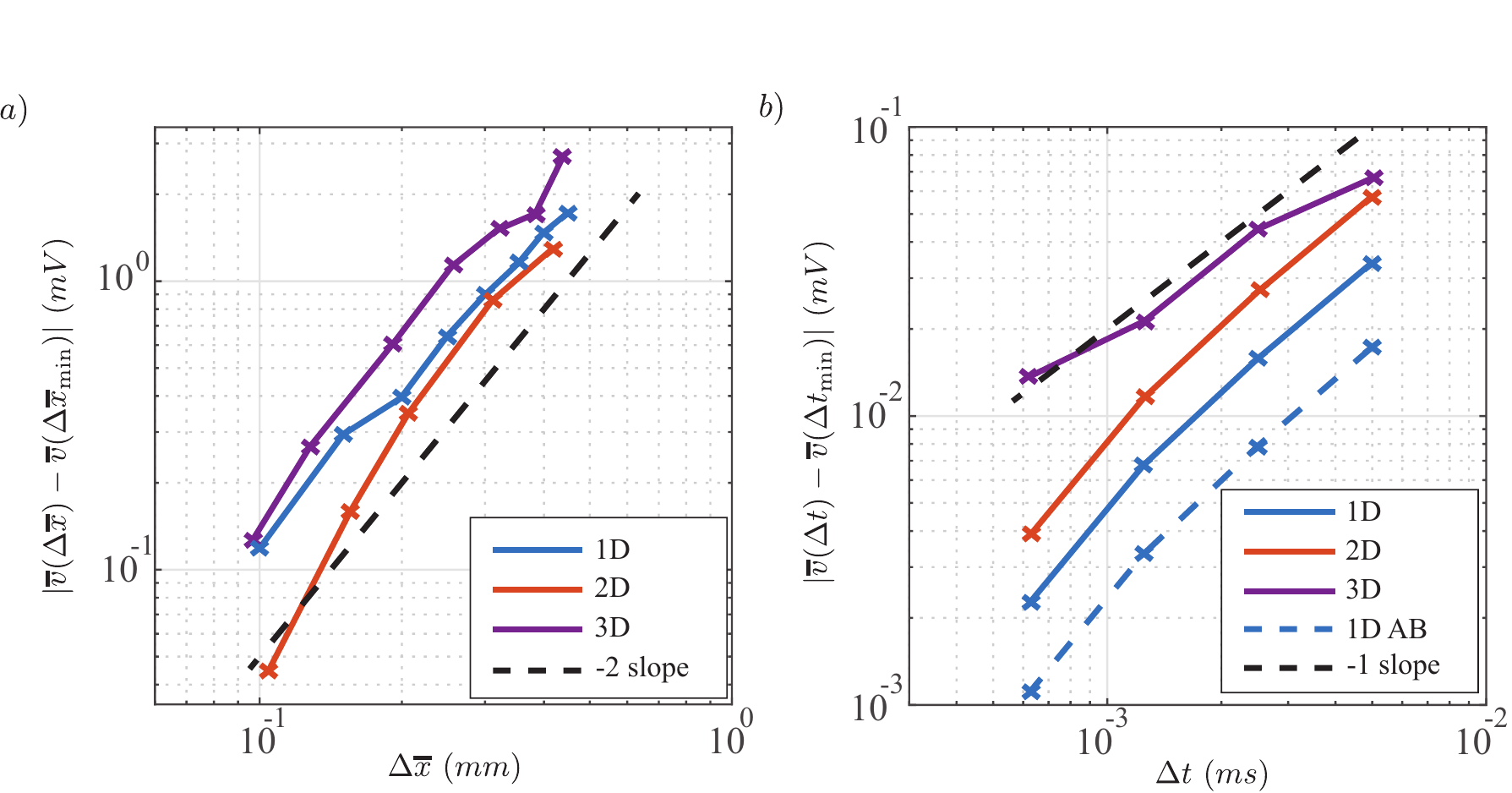}}
\caption{\textcolor{black}{(a) Spatial and (b) temporal numerical accuracy of the 1D, 2D, and 3D monodomain solver. In (a) the error on the average transmembrane potential (computed at time $t=50$~ms for $\Delta t = 0.005$~ms) is shown as a function of the mean grid size, $\overline{\Delta x}$. In (b) the same quantity (at time $t=50$~ms for $\overline{\Delta x} = 0.31$~mm) is reported as a function of the timestep, $\Delta t$. 
}}
\label{fig:accuracy}
\end{figure}
\\ \indent
\textcolor{black}{
The corresponding numerical accuracy of the solver is reported in Figure~\ref{fig:accuracy} in the case of the 1D, 2D and 3D monodomain model. Figure~\ref{fig:accuracy}(a) indicates the second order spatial accuracy of the FV method by showing the error on the average transmembrane potential over the domain at time $t=50$~ms (and timestep set to $\Delta t = 0.005$~ms) as a function of the average grid size, $\overline{\Delta x}$.
On the other hand, Figure~\ref{fig:accuracy}(b) reports the same quantity (evaluated with $\overline{\Delta x}=0.31$~mm) as a function of the timestep $\Delta t$, thus retrieving the first order temporal accuracy of the Rush--Larsen temporal integration scheme, where the non--gating variables are solved using forward Euler. The numerical error could be reduced by adopting a second--order Adams--Bashforth scheme for the non--gating variables (dashed blue line for the 1D case), although a modified second order Rush--Larsen scheme \cite{perego2009efficient} would be needed to attain a second order accuracy.}  
\begin{table}[h!!]
\centering
\begin{tabular}{c|c c c | c c c}
\vtop{\hbox{\strut  ~~\bf Grid }\vspace{-0.2cm}\hbox{\strut $\Delta x$, cells}} &\vtop{\hbox{\strut  ~~~\bf CPU }\vspace{-0.2cm}\hbox{\strut bidomain}}&\vtop{\hbox{\strut  ~~~\bf GPU }\vspace{-0.2cm}\hbox{\strut bidomain}} &\vtop{\hbox{\strut \bf speedup }\vspace{-0.2cm}\hbox{\strut bidomain}} &\vtop{\hbox{\strut  ~\bf CPU }\vspace{-0.2cm}\hbox{\strut monod.}}&\vtop{\hbox{\strut  ~\bf GPU }\vspace{-0.2cm}\hbox{\strut monod.}} &\vtop{\hbox{\strut \bf speedup }\vspace{-0.2cm}\hbox{\strut ~monod.}}\\
\hline
\vtop{\hbox{\strut  0.5~mm, }\vspace{-0.2cm}\hbox{\strut ~20'160}} & 0.027~s & 0.0015~s & 18 & 0.018~s & 0.0002~s  & 90 \\
\vtop{\hbox{\strut  0.2~mm, }\vspace{-0.2cm}\hbox{\strut 315'000}} & 0.45~s & 0.017~s & 27 & 0.20~s & 0.0013~s & 154 \\
\vtop{\hbox{\strut  ~0.1~mm, }\vspace{-0.2cm}\hbox{\strut 2'520'000}} & 3.7~s & 0.09~s & 41 & 1.6~s & 0.009~s & 177 \\
\end{tabular}
\caption{\textcolor{black}{Wall--clock time for integrating a single bidomain and monodomain time step for the three Cartesian grids considered in Figures~\ref{fig:MatlabMatlab}--\ref{fig:MatlabPython}. The CPU time is obtained using a single core Intel(R) Xeon(R) Gold 6230 with 2.10GHz, whereas the GPU time corresponds to a Tesla V100 from Nvidia.} }
\label{tab:speedup}
\end{table}
\\ \indent 
\textcolor{black}{
Table~\ref{tab:speedup} reports the wall--clock time for solving a single bidomain and monodomain timestep on the benchmark Cartesian domain using a single CPU core or GPU device. Running the GPU version of the code yields a significant speedup in all cases, which increases as the grid gets more refined owing to the better balance of workload across the GPU threads running in parallel. It can be observed, that the speedup is larger in the case of monodomain computations with respect to the bidomain counterpart because the Arnoldi iteration and the solution of the corresponding Hessenberg system in the GMRES algorithm are based on the Lapack library running on the CPU.}
\\ \indent
The convergence of the numerical solution over the cardiac domains used in sections~\ref{sec:domain}~and~\ref{sec:results} for the electrical conductivities reported in Table~\ref{tab:conductivities} is assessed by monitoring the average transmembrane potential as a function of time, $\overline{v}(t)$. 
Figure~\ref{fig:ValidationRealGeometry}(a) shows $\overline{v}(t)$ solved over the 1D network of fast conduction bundles using the monodomain model coupled with the Courtemanche cellular model with a timestep equal to $\Delta t = 0.001$ ms, thus showing good convergence for a spatial discretization finer than $\Delta x = 0.25$ mm (corresponding to a number of cells equal to $2'000$). 
On the other hand, the monodomain equations over the right Purkinje 2D media coupled with the Stewart cellular model are at convergence for the number of triangles of about 54'000 (i.e. $\Delta x = 0.45$ mm). 
\textcolor{black}{
The bidomain solution over the 3D left atrium becomes grid independent for a number of cells around 5'500'000 ($\Delta x = 0.31$ mm), but a coarser grid with 1'500'000 cells (corresponding to $\Delta x = 0.53$ mm) yields a slower propagation of the depolarization front corresponding to a time delay of the electrical activation of the chamber  below 3.5$\%$ the heart beating period (equal to 750~ms for a heart rate of 80~b.p.m.). Based on these results, a 1D grid made of 2'500 linear elements and a 2D grid of 108'000 triangles have been used to discretize the fast conduction and the Purkinje networks with a timestep of $0.0001$~ms.}
\textcolor{black}{
Hence, the electrophysiology of the 3D myocardium is integrated with a timestep of $\Delta t=0.005$~ms using using 1'500'000, 2'500'000 and 5'500'000 tetrahedra for the left atrium, right atrium and ventricles, respectively. The electrophysiology of each 3D chamber is integrated using a dedicated GPU card Tesla V100 from Nvidia and the wall--clock time to run a single heart beat is thus given by that of solving the ventricular electrophysiology, which is equal to 7.9~hours (corresponding to a speedup of 60 times with respect to the serial CPU version of the code). Remarkably, since the computational cost is dominated by the 3D solution of the bidomain equations, it can be greatly reduced by using a monodomain approach as it avoids solving a large linear system for $v^{ext}$, thus obtaining a wall--clock time of 1.4~hours per heart beat.}

\section{Results: electrophysiology of the whole heart}\label{sec:results}
\begin{figure}[h!]
\centering
\includegraphics[width=1\textwidth]{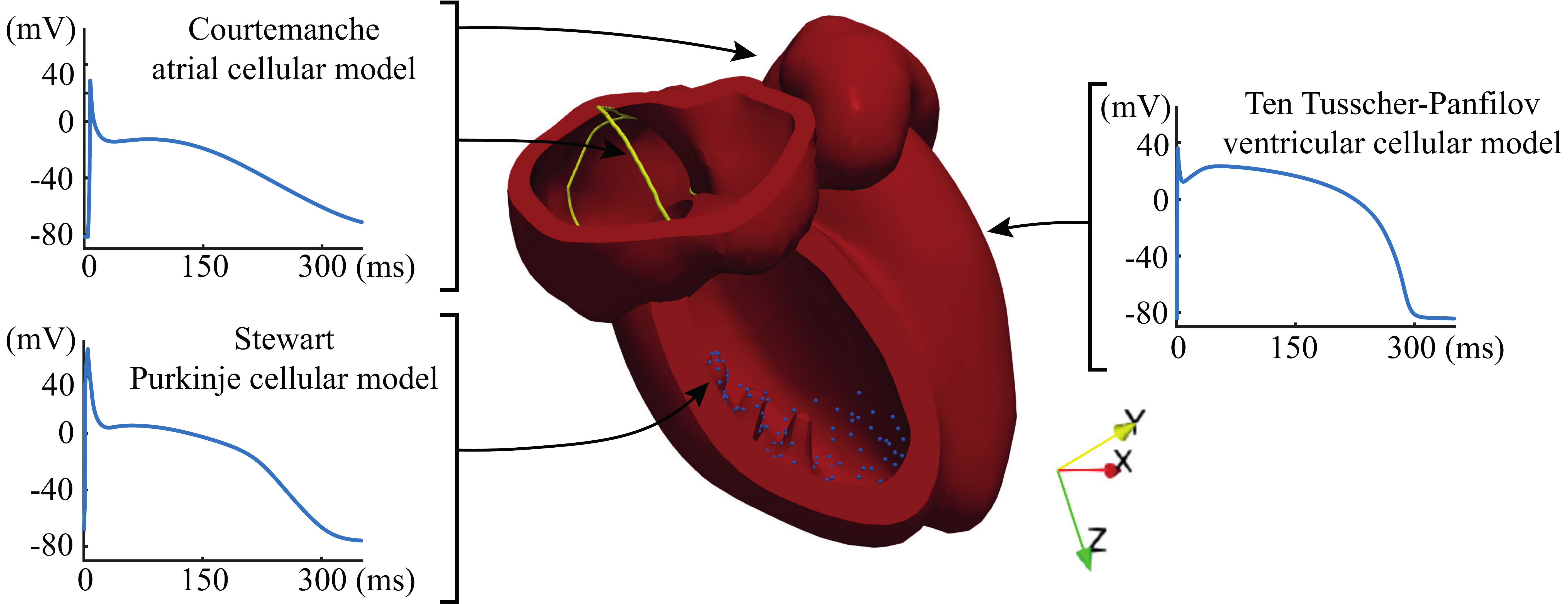}
\caption{Action potential at different cardiac locations. a) The Courtemanche cellular model \cite{courtemanche1998ionic} is used in the 1D atrial bundles and in the 3D atrial myocardium, b) the Stewart  model \cite{stewart2009mathematical} is adopted for the Purkinje network and c) the ten~Tusscher-Panfilov  model \cite{ten2006} governs the action potential in the 3D ventricular myocardium.}
\label{fig:DifferentPotential}
\end{figure}

The electrophysiology of the whole heart is solved using the cardiac configuration introduced in \S~\ref{sec:domain} (Figure~\ref{fig:subdivided}),
 which is composed of a 1D network of bundles, a 2D surface to mimic the Purkinje placed at the ventricular endocardium and 3D media for atrial and ventricular myocardium.  In order to better account for their heterogeneous electrophysiology properties three different cellular models are adopted (Figure~\ref{fig:DifferentPotential}). 
 In particular,   the Courtemanche model \cite{courtemanche1998ionic} is used for the atrial bundles and myocardium, which  has a resting potential of $-80$~mV and is characterized by rapid repolarization occurring in about 200~ms. On the other hand, the high peak of depolarization followed by a stable plateau phase of about 250 ms observed in the Purkinje cells is modelled through the Stewart cellular model \cite{stewart2009mathematical}, whereas the ionic fluxes across the ventricular myocytes  are governed by the ten~Tusscher--Panfilov cellular model \cite{ten2006} exhibiting a resting potential of about $-85$~mV
and a longer depolarization plateau of about 300~ms which is related to a longer muscular contraction of the fibers.
Furthermore, the depolarization front propagates at different velocity through these media according to different electrical conductivities, which have been set in the electrophysiology model as summarized in Table~\ref{tab:conductivities}. Since the monodomain and bidomain models are equivalent in the case of a 1D domain, only a single electrical conductivity has been set so to reproduce the propagation velocity  reported in the literature \cite{hall2015guyton}.
Owing to the high density of the Purkinje fibers and their heterogeneous intracellular and extracellular orientation, the Purkinje network is modelled as a uniform media governed by the monodomain equation with an isotropic conductivity tensor with its components set to reproduce a propagation velocity of $4$~m/s \cite{del2021electrophysiology,dossel2012computational,harrild2000computer}.
The intracelluar and extracellular electrical heterogeneity in the 3D ventricular myocardium is accounted by setting an anisotropic  conductivity tensor in the bidomain equations as reported in the literature \cite{niederer2011verification} (Table~\ref{tab:conductivities}). Owing to the lack of data on the atrial myocardium conductivity, the same conductivity tensors as in the ventricles have been used but rescaled by a factor to match the propagation velocity measured experimentally.
\setlength{\tabcolsep}{3pt}
\begin{table}[t!]
\begin{tabular}{p{1.5cm} p{2cm}p{2cm}p{1.5cm}}
\vtop{\hbox{\strut ~~~ \bf{heart}  }\vspace{-0.2cm}\hbox{\strut \bf{component}}}
 &  \vtop{\hbox{\strut  ~~\bf{cell/PDE}  }\vspace{-0.2cm}\hbox{\strut ~~~~ \bf{model}}} 
 & 
 \vtop{\hbox{\strut  \bf{conductivity}  }\vspace{-0.2cm}\hbox{\strut ~ \bf{(mS/mm)}}} 
 & \vtop{\hbox{\strut ~~~ ~\bf{reference}  }}                                                                                 \\ \hline
  \vtop{\hbox{\strut 1D internodal   }\vspace{-0.2cm}\hbox{\strut bundles }} &
  \vtop{\hbox{\strut Courtemanche,   }\vspace{-0.2cm}\hbox{\strut isotropic } \vspace{-0.2cm}\hbox{\strut monodomain }}
& \vtop{\hbox{\strut $m_\parallel = 1.29 $   }}                                                                   & 
\vtop{\hbox{\strut \textcolor{black}{corresponding to}  }\vspace{-0.2cm}\hbox{\strut a velocity 1.54 m/s  } \vspace{-0.2cm}\hbox{\strut \cite{harrild2000computer,dossel2012computational,del2021electrophysiology} }}
               \\ \hline
 \vtop{\hbox{\strut 2D Purkinje   }\vspace{-0.2cm}\hbox{\strut network }} &   
  \vtop{\hbox{\strut Stewart, isotropic  }\vspace{-0.2cm}\hbox{\strut monodomain }}
  & \vtop{\hbox{\strut $m_\parallel = m_{/}  = 3.95 $  }}   
  &   
  \vtop{\hbox{\strut \textcolor{black}{corresponding to}  }\vspace{-0.2cm}\hbox{\strut  a velocity 4.0 m/s  \cite{hall2015guyton} } }
   \\ \hline
\vtop{\hbox{\strut 3D ventricles     }}                 & 
  \vtop{\hbox{\strut ten~Tusscher--  }\vspace{-0.2cm}\hbox{\strut Panfilov, } \vspace{-0.2cm}\hbox{\strut bidomain } }
&
  \vtop{\hbox{\strut $m^{ext}_\parallel=0.62$  }\vspace{-0.15cm}\hbox{\strut $m^{ext}_\perp=m^{ext}_{/}=0.24$} \vspace{-0.15cm}\hbox{\strut $m^{int}_\parallel=0.17$}\vspace{-0.15cm}\hbox{\strut $m^{int}_\perp=m^{int}_{/}=0.019$  }}
  & 
  \vtop{\hbox{\strut data from \cite{niederer2011verification}           }}   
                                                                                      \\ \hline
\vtop{\hbox{\strut 3D atria     }}                     & 
 \vtop{\hbox{\strut Courtemanche,   }\vspace{-0.2cm}\hbox{\strut bidomain } }
       & 
  \vtop{\hbox{\strut same as ventricles   }\vspace{-0.2cm}\hbox{\strut but rescaled by} \vspace{-0.2cm}\hbox{\strut a factor 1.05 }} &
  \vtop{\hbox{\strut \textcolor{black}{corresponding to}   }\vspace{-0.2cm}\hbox{\strut a longitudinal velocity} \vspace{-0.2cm}\hbox{\strut  0.5 m/s \cite{hall2015guyton,harrild2000computer}  }}
\end{tabular}
  
\caption{\textcolor{black}{Monodomain and bidomain electrical conductivities of the various cardiac components, as defined in section~\ref{sec:num_meth}. }}
\label{tab:conductivities}
\end{table} 
\begin{figure}[h!]
\centering
\includegraphics[width=1\textwidth]{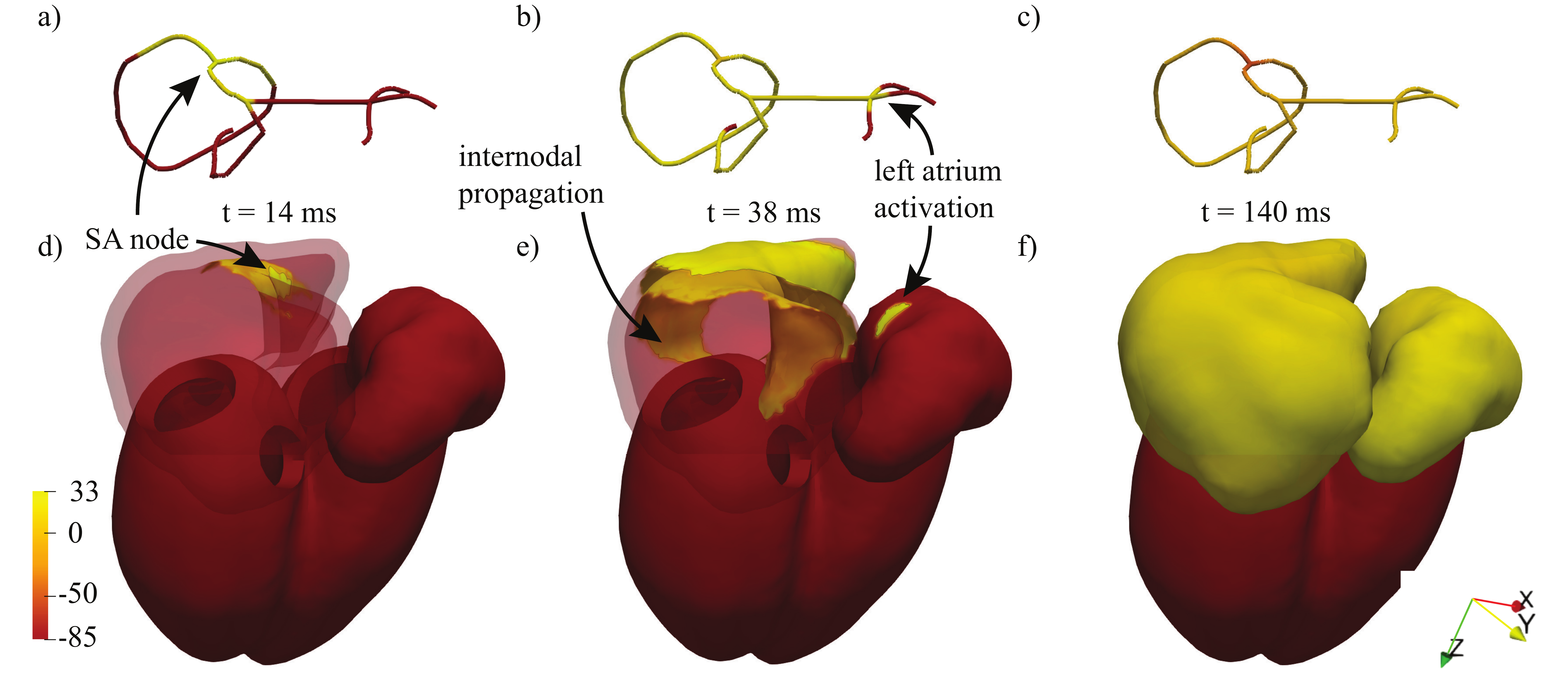}
\caption{
Atrial depolarization. Transmembrane potential in the fast conduction atrial bundles at a) $t=14$~ms b) $t=38$~ms and c) $t=140$~ms, whereas the corresponding depolarization of the 3D myocardium is shown in d), e) and f), respectively.
In (a,d) the depolarization front has just been initiated in the SA--node,  in (b,e)  most of 3D right atrium is depolarized and the depolarization front just reached the left atrium through the 1D bundles. In c) both atria are depolarized and the right one starts the repolarize.
}
\label{fig:stepsAtria}
\end{figure}
\subsection{Healthy electrophysiology}
We can now analyze the whole cardiac electrical activation in a healthy heart.
Figure~\ref{fig:stepsAtria} shows the depolarization of the fast conduction atrial bundles (panels a,b,c) and of the 3D atrial myocardium (panels d,e,f) at three different time instants, with $t$ defined as the time lag with respect to the SA--node activation (corresponding to $t=0$).
In both 1D and 3D media, the transmembrane potential, $v$, has an initial resting value of $-80$~mV  (red isolevel) and transiently reaches a positive value (yellow isolevel) as the depolarization front advances.
The latter originates at the SA--node, where the effect of the pace--maker cells translates into an initial electrical stimulus~\eqref{eq:stimulus} activating the SA--node (Figure~\ref{fig:stepsAtria} a), which  then advances simultaneously into the three internodal pathways, namely the Thorel's posterior internodal tract, the Wenckebach's bundle-middle internodal tract and the anterior internodal tract that further bifurcates in the Bachmann's bundle towards the left atrium (Figure~\ref{fig:stepsAtria} b).
The propagation fronts in these three internodal pathways then recollect at the bottom of the atrial network into the AV--node (see Figure~\ref{fig:stepsAtria} b) after about 25-40 ms from the SA--node activation, first throughout  the anterior and middle bundle and later throughout the posterior one.
%
In the meantime the depolarization front propagates in the 1D network of bundles, it activates the surrounding atrial myocardium through the CNs, thus triggering another depolarization front in the 3D media, as shown by the incipient myocardial activation near by the  SA--node in  Figure~\ref{fig:stepsAtria}(d).
Although the conduction speed in the 3D myocardium is anisotropic and the transmembrane depolarization advances faster in the directions aligned with the muscular fibers (owing to a larger electrical conductivity), the myocardium depolarization is about three times slower than the one in the bundles. This leads to a complete activation of both 3D atrial chambers after about 140~ms, as visible in Figure~\ref{fig:stepsAtria} (f). 
Interestingly, the endocardial and epicardial depolarization fronts in Figure~\ref{fig:stepsAtria}~(e), reveal that  
the epicardium depolarizes with few milliseconds delay with respect to the endocadium, which corresponds to the time lag needed by the 3D depolarization front (originated at the bundles placed within the  endocardium) to propagate across the atrial wall in the cross--fiber directions.
\begin{figure}[!h]
\centering
\includegraphics[width=1\textwidth]{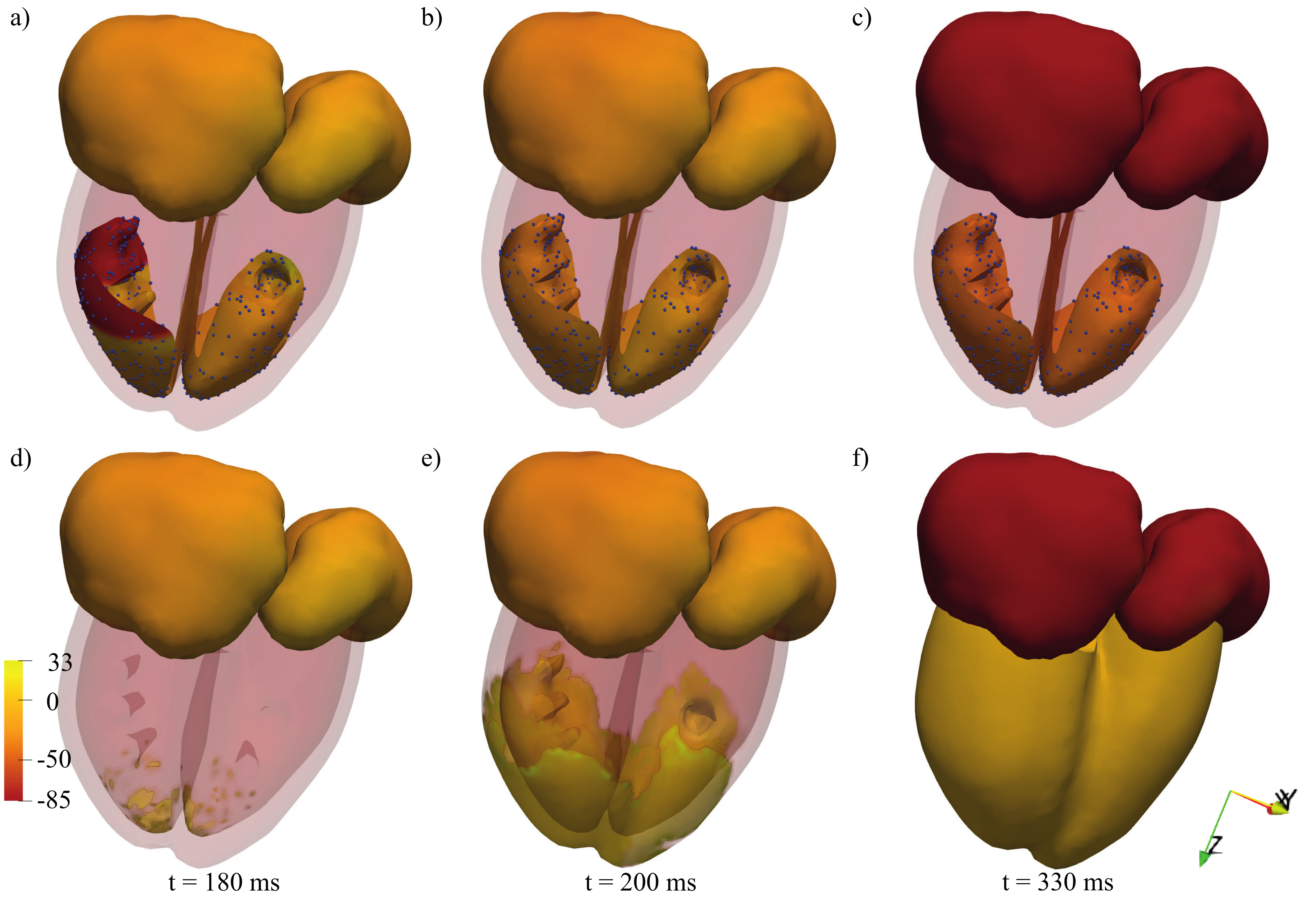}        
\caption{Ventricular depolarization. Transmembrane potential at a) $t=180$~ms b) $t=200$~ms and c) $t=330$~ms showing the action potential propagation front in the Purkinje and the locations of the PMJs coupling the Purkinje with the ventricular myocardium. The same data are reported in panels d) e) and f) but using a different transparency so to better visualize the 3D ventricular depolarization. In panels (a,b) both atria have been activated, whereas in panel (c) they are completely depolarized.}
\label{fig:3DFullTimeSteps}
\end{figure}
\\ \indent In non--pathological profiles, as is the case here, the signal carried into the AV--node by the internodal pathways propagates from the left atrium to the ventricles only through the AV--node itself. In the AV--node, the propagation speed of the depolarization front greatly reduces, yielding a delay of about 100~ms between the atrial and the subsequent ventricular depolarization (Figure~\ref{fig:3DFullTimeSteps}~a).
After the depolarization front swept the AV--node, it reaches the bundle of His before  propagating in the two ventricular chambers through  left bundle and right bundles (Figure~\ref{fig:3DFullTimeSteps} a), which, in turn, are electrically connected to the Purkinje network. The latter carry the depolarization front in both ventricular chambers with a propagation speed of about 4 m/s (roughly ten times  the surrounding myocardial tissue), first activating the lower part of the ventricle (Figure~\ref{fig:3DFullTimeSteps} d,e) and then the upper part (Figure~\ref{fig:3DFullTimeSteps} e,f). As visible in the same panels, the activation of the 2D Purkinje network precedes the surrounding 3D ventricular activation, thus yielding a more synchronous depolarization of the 3D media.  
As visible in the upper panels of Figure~\ref{fig:3DFullTimeSteps}, the 3D myocardium is electrically activated by the 2D Purkinje network through the PMJs with an orthodromic delay of 5 ms. 
Importantly, when the ventricles are almost completely depolarized, the atria are  repolarizing (Figure~\ref{fig:3DFullTimeSteps} e) and, successively, when the ventricles are completely activated, the atria are fully repolarized (Figure~\ref{fig:3DFullTimeSteps} f).
%
%
\begin{figure}[!htbp]
    \centering
    \includegraphics[width=.9\textwidth]{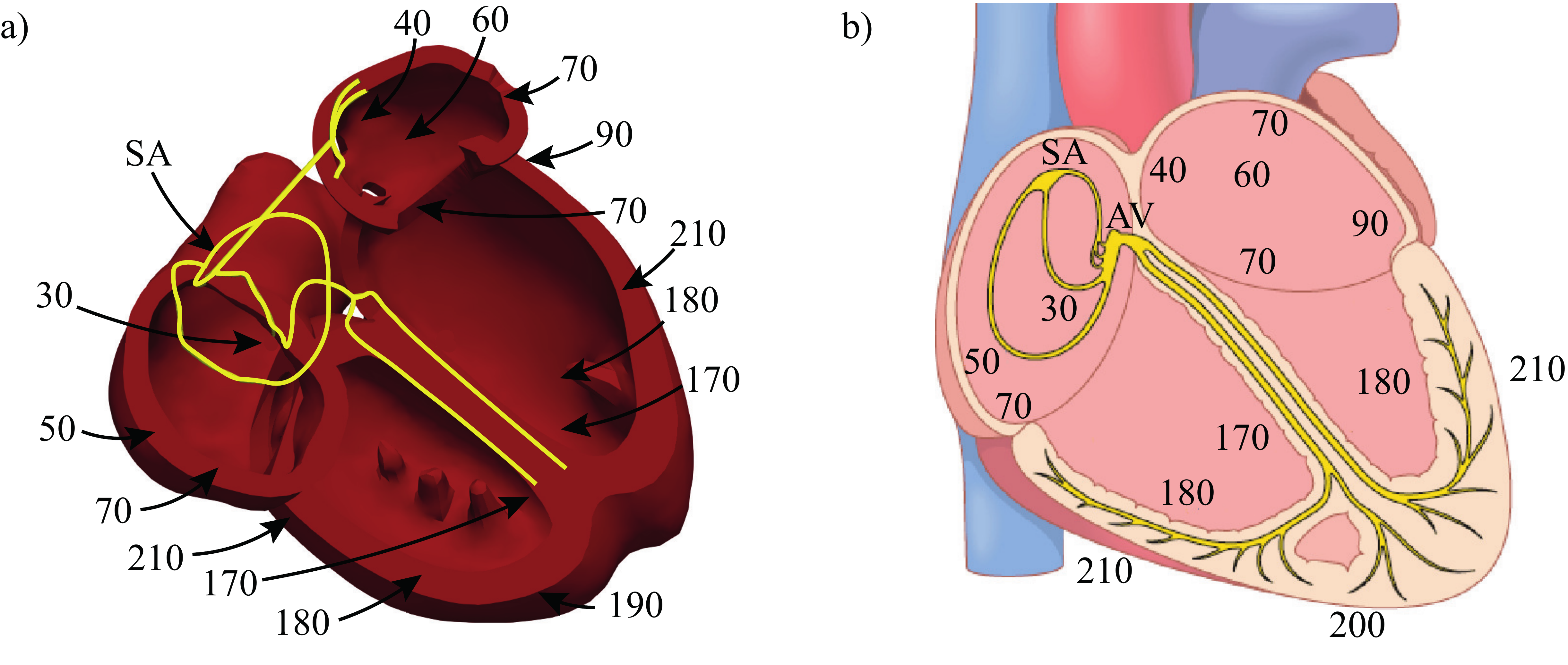}
    \caption{Time lag (in $ms$) of the cardiac depolarization at various heart locations with respect the SA--node stimulus according to a) our numerical model  and b) medical atlas \cite{hall2015guyton}.}
    \label{fig:PhysicalComparison}
\end{figure}
\\ \indent  Figure~\ref{fig:PhysicalComparison} compares the electrical activation at various cardiac locations reported in medical atlas \cite{hall2015guyton} with respect to those obtained by our computational model, where each number corresponds to the  time interval in milliseconds that lapse between the activation of SA--node and that of the location indicated by the number: an overall good agreement can be observed.
In particular, the fast atrial 1D conduction system (internodal pathways) correctly ensures the activation of the atrioventricular node after 30 milliseconds from the activation of the SA--node, with a perfect match with what observed in--vivo reality. Furthermore, the slower conduction velocity in the AV--node and the subsequent rapid spreading of the depolarization front in the 2D Purkinje network provide the correct activation of the entire ventricular endocardium, including the papillary muscles.
%
%
%
%
\subsection{Pathologic and aided electrophysiology: bundle branch block and artificial pacing} \label{eq:patho}
\begin{figure}[!h]
\centering
\makebox[\textwidth][c]{\includegraphics[width=1.2\textwidth]{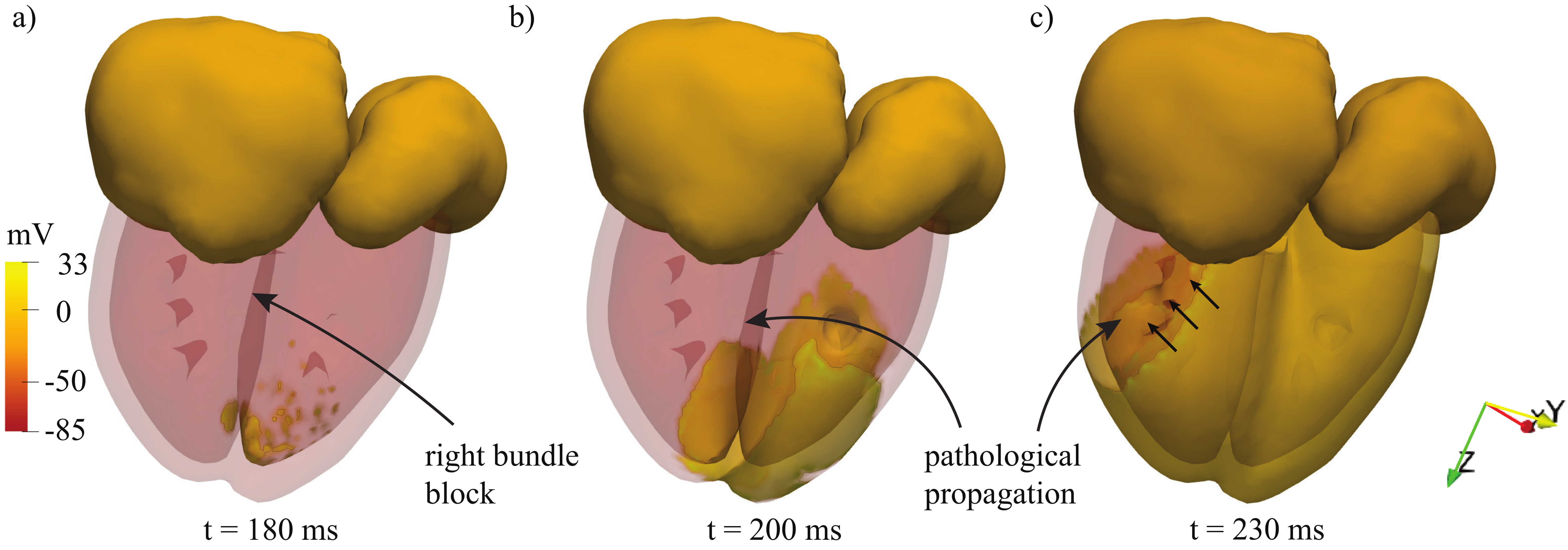}}
\caption{Pathologic ventricular activation in the case of right bundle block at various time instants with respect the SA--node activation. In particular, the time instants of panels (a) and (b) corresponds to the ones for the  healthy cases reported in Figure~\ref{fig:3DFullTimeSteps}. The black arrow in panel c) highlights the delayed right ventricle depolarization caused by the disease.}
\label{fig:BlockRight}
\end{figure}

\begin{figure}[!h]
\centering
\makebox[\textwidth][c]{\includegraphics[width=1.2\textwidth]{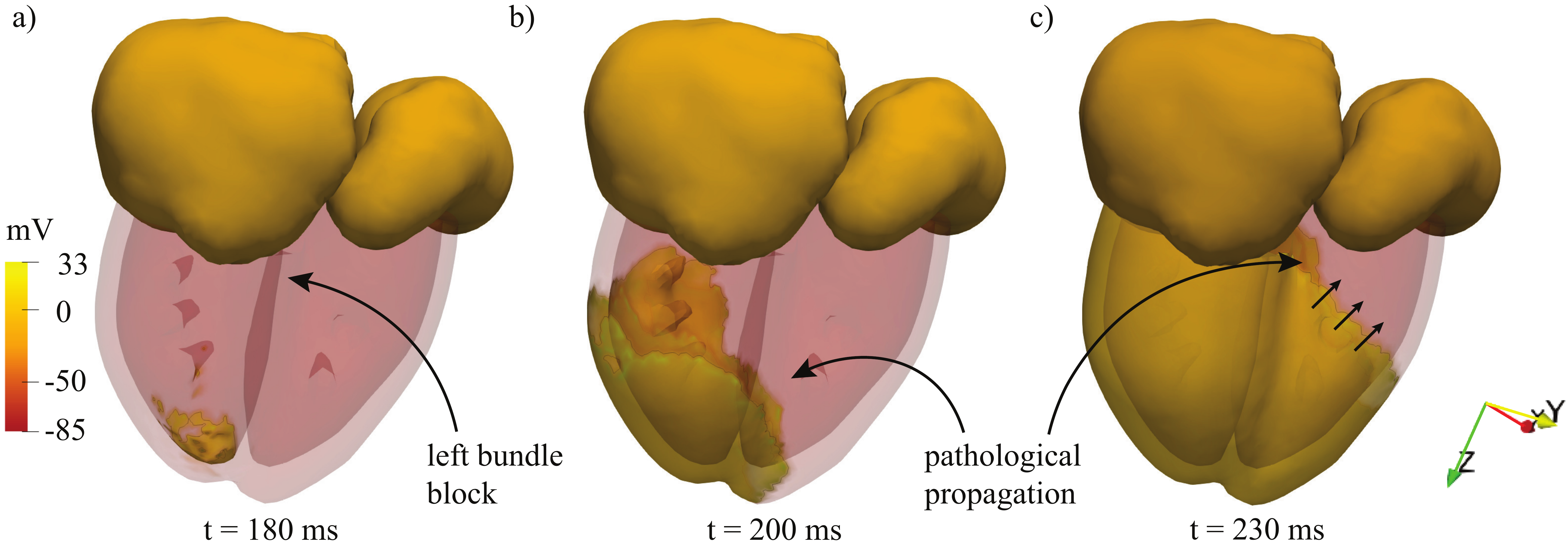}}
\caption{Pathologic ventricular activation in the case of left bundle block at various time instants with respect the SA--node activation. In particular, the time instants of panels (a) and (b) correspond to the ones for the  healthy cases reported in Figure~\ref{fig:3DFullTimeSteps}. The black arrow in panel c) highlights the delayed left ventricle depolarization caused by the disease. }
\label{fig:BlockLeft}
\end{figure}

The present high--fidelity computational framework for the whole heart electrophysiology allows also to model cardiac pathologies and predict the effect of medical devices, such as the artificial pacing applied to a patient affected by a bundle branch block.
\begin{figure}[!t]
\centering
\makebox[\textwidth][c]{\includegraphics[width=1.1\textwidth]{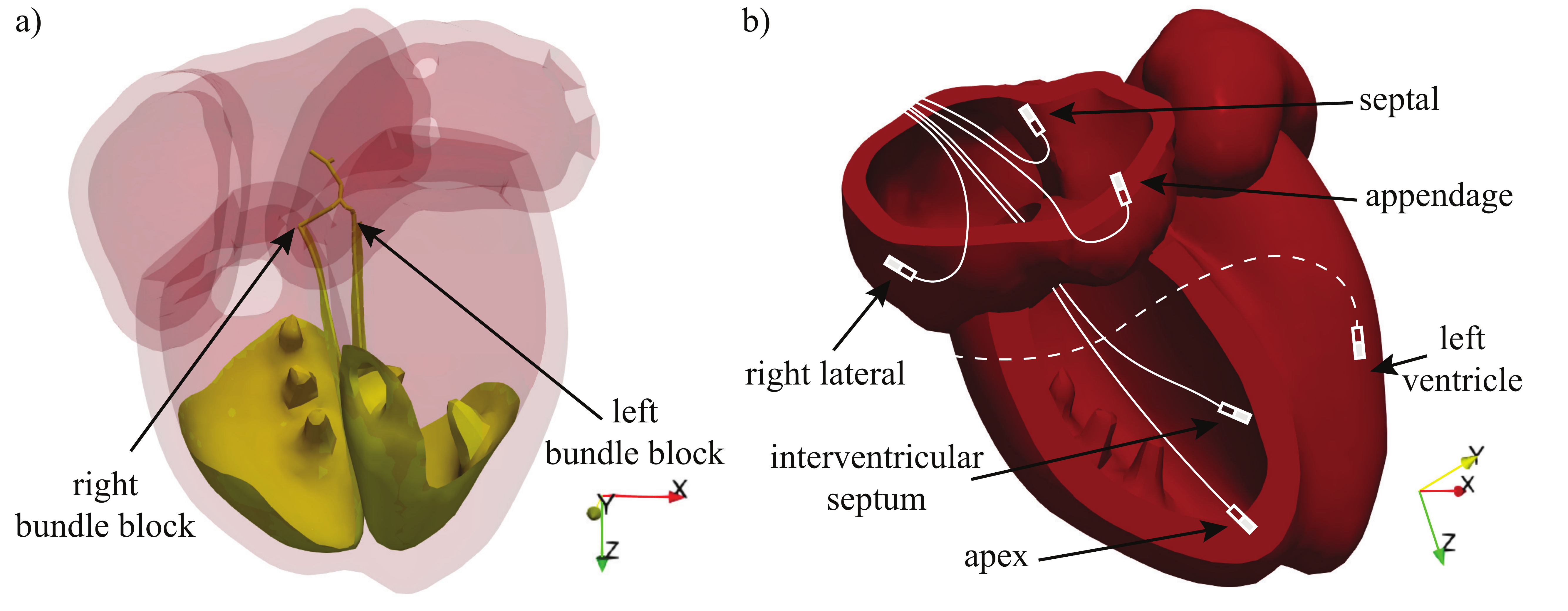}}
\caption{a) Positions of the right and left ventricular bundle blocks. b) Most common leads potitioning for atrial and ventricular pacemaking \cite{pakarinen2013minimizing}.}
\label{fig:FullPat}
\end{figure}
%
%
The latter consists of the delay or  blockage of the electrical propagation within a ventricular bundle (see Figure~\ref{fig:FullPat}a), thus implying a delayed activation of some areas of the ventricular myocardium and a consequent anomaly in the activation/contraction profile of the ventricle \cite{takeshita1974effect,xiao1991effect}.
Possible causes originating a bundle branch block include heart attacks (myocardial infarction involving the bundles), myocarditis (viral/bacterial infection of the heart muscle), cardiomyopathy (thickened/stiffened or weakened heart muscle), congenital heart abnormality (such as atrial septal defect) or even high blood pressure (hypertension) \cite{mahrholdt2006clinical,maron2006contemporary}. 
\\ \indent
The occurrence of this pathology is included in the 1D fast conductivity bundles and in the 2D Purkinje network as a local reduction of the electrical conductivities proportionally to the severity of the bundle branch delay, whereas a null conductivity tensor is used to simulate a complete block of the bundle.
The resulting pathologic  activation in the case of a right bundle branch block is reported in Figure~\ref{fig:BlockRight} showing that, despite the depolarization of the left ventricular myocardium is correctly initiated after about 180 ms (panel \ref{fig:BlockRight}a),
the missing propagation of the depolarization front through the right Purkinje network prevents the normal depolarization of the ventricular myocardium observed in the healthy cases (Figure~\ref{fig:3DFullTimeSteps} d,e).
However, as the left and right ventricular myocardium are a unique 3D excitable media, the right ventricular depolarization is triggered by the surrounding left ventricular one with a delay of about $20$~ms and the propagation front then travels throughout the chambers (panel~\ref{fig:BlockRight}b). Nevertheless, as the conduction speed in the 3D myocardium is about ten times slower than the one in the Purkinje network, the depolarization of the right chamber results significantly delayed with respect to healthy conditions with an asynchronous depolarization of the apical, equatorial and basal myocytes (Figure~\ref{fig:BlockRight}c), which would entail an inefficient systolic contraction \cite{hall2015guyton}.
Vice--versa, the presence of a left bundle block, shown in Figure~\ref{fig:BlockLeft}, yields a delayed activation of the left ventricle owing to the missing propagation in the left ventricular bundle and Purkinje network. 
\begin{figure}[!h]
\centering
\makebox[\textwidth][c]{\includegraphics[width=1\textwidth]{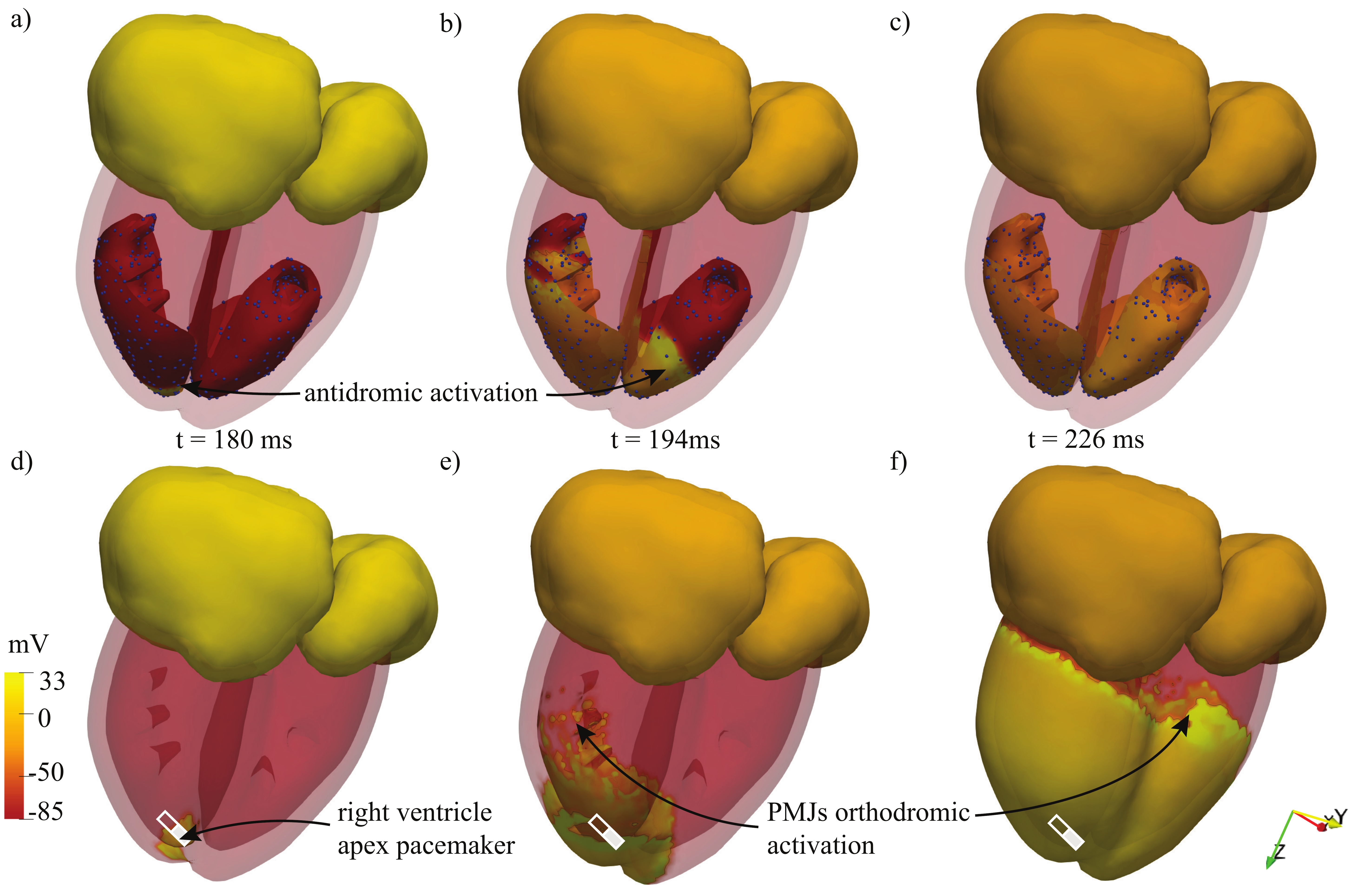}}
\caption{
Cardiac electrophysiology with an artificial pacing at the ventricular apex, the white symbol indicates the lead location. Panels (a,b,c) show the action potential front in the Purkinje network while panels (d,e,f) describe corresponding depolarization of the surrounding 3D myocardium. The AV--node communication is interrupted (AV block).
}
\label{fig:VentricularPacing}
\end{figure}
\\ \indent 
Bundle block pathologies, as well as other cardiac diseases such as sinus node dysfunction and intermittent AV block \cite{strauss1976electrophysiologic,gillis2006reducing} are often treated with artificial stimulation through the implantation of an artificial cardiac pacemaker \cite{pakarinen2013minimizing}, which consists of inserting an artificial lead in contact with the internal muscular wall (endocardium) inducing the periodic depolarization of the surrounding tissue. The effect of an implanted pacemaker lead can be accounted for in the model through an additional stimulation current $I^s$ in equation~\eqref{eq:electro} acting on the 3D myocardium and localized at the lead position. Among the most common atrial (septal, right lateral, appendage) and ventricular (apex, interventricular septum, left ventricle) leads implantation sites reported in Figure~\ref{fig:FullPat}(b), here we consider a ventricular apex pacing to mitigate a pathologic atrioventricular block (inability of the signal to cross the AV--node) simulated by setting the conductivity of the AV--node to zero. 
As shown in Figure~\ref{fig:VentricularPacing}(a), although the ventricular myocardium is not activated by the fast conduction Purkinje complex, as in healthy conditions, the pacing lead implanted within the apical tissue of the right ventricle induces an electrical stimulus with a delay of 160 seconds with respect to the SA--node. 
\begin{figure}[!t]
\centering
\makebox[\textwidth][c]{\includegraphics[width=1\textwidth]{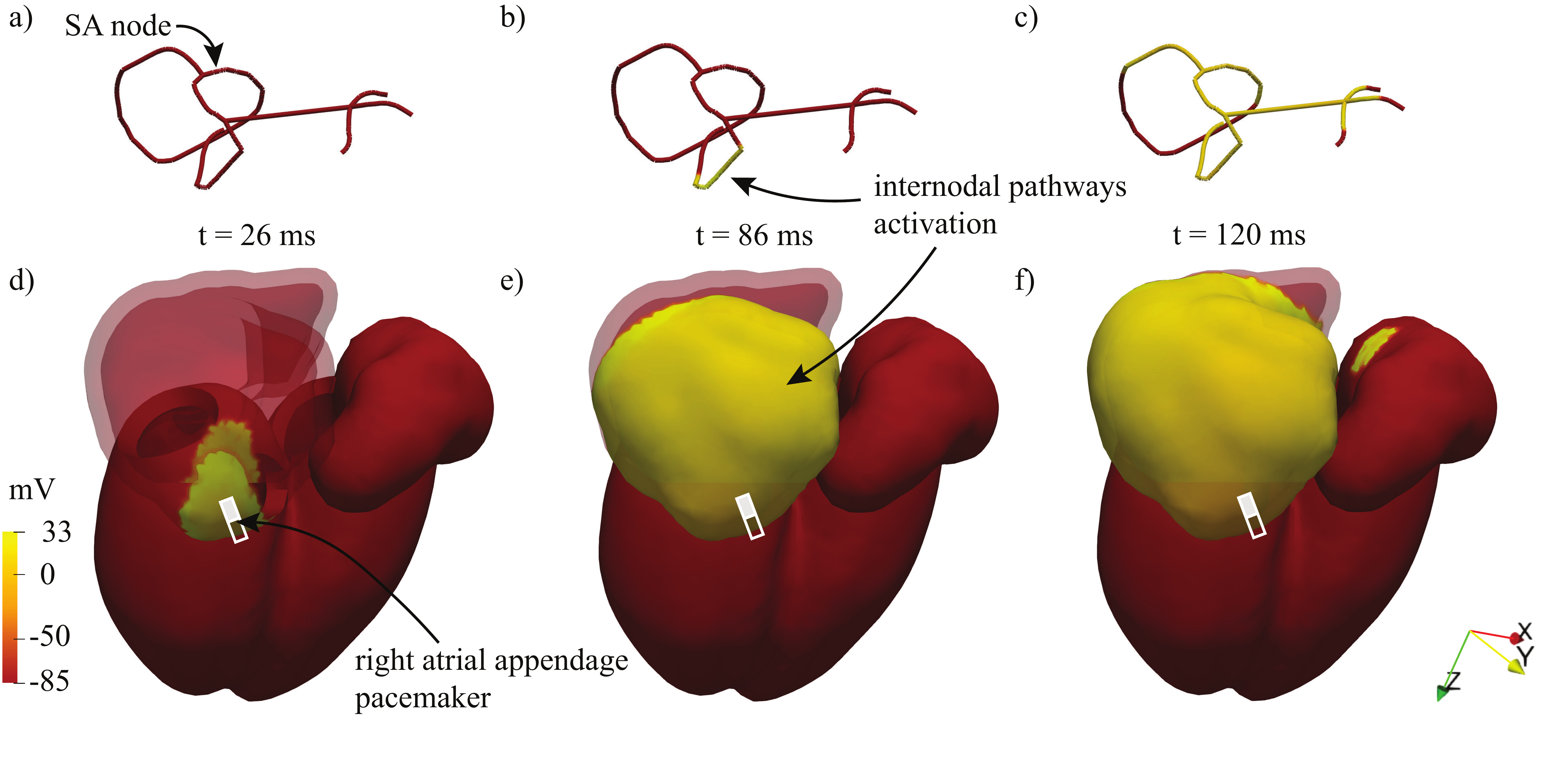}}
\caption{Cardiac electrophysiology with an artificial pacing at the atrial appendage, the white symbol indicates the lead location. Panels (a,b,c) show the action potential depolarization front within the internodal pathways, whereas  panels (d,e,f) indicate the corresponding 3D myocardium. 
}
\label{fig:AtrialPacing}
\end{figure}
In particular, the depolarization front propagates from the 3D myocardium activated by the lead to the Purkinje network through the PMJs (with an antidromic delay of 3 ms, see Figure~\ref{fig:VentricularPacing} b,e), and then rapidly propagates through the rest of the Purkinje network which, in turn, triggers the depolarization of the underlying 3D  myocardial tissue through the downstream PMJs (with an  orthodromic communication delay, see Figure~\ref{fig:VentricularPacing} e). Hence, in agreement with the medical evidence, the presence of an artificial stimulation through an implanted lead manages to activate the Purkinje network downstream the bundle block and to recover an effective depolarization of the ventricular myocardium in a similar fashion as the healthy depolarization pattern studied above and reported in Figure~\ref{fig:3DFullTimeSteps}. 
\\ \indent In addition, Figure~\ref{fig:AtrialPacing} shows the cardiac electrophysiology corresponding to an atrial pacemaking, where the lead is positioned in the right atrial appendage \cite{cox1996operative}, one of the most frequent implantation sites for an atrial lead \cite{pakarinen2013minimizing}. Initially, the electrical stimulus provided by the lead only depolarizes a surrounding piece of the 3D atrial myocardium (panel~\ref{fig:AtrialPacing}~d), whereas the atrial 1D bundles are not directly activated by the lead (panel~\ref{fig:AtrialPacing}~a). The electrical depolarization front propagates in the 3D myocardium until reaching and, consequently, activating the fast conduction bundles (see panel~\ref{fig:AtrialPacing}~b), which then rapidly propagates the depolarization front in the whole 1D network, including the left atrial network (panel~\ref{fig:AtrialPacing}~c). 
The combined effect of slow (3D) and fast (1D) depolarization fronts induced by the atrial lead thus yields a homogeneous activation of the left atrium. Nevertheless, compared to healthy propagation shown in Figure~\ref{fig:stepsAtria}~b), the atrial depolarization occurs with a delay of approximately 95 ms.

\section{Discussion}  \label{sec:discussion}
In this work, a numerical framework for solving the cardiac electrophysiology of the whole human heart in healthy and pathologic conditions has been proposed. According to the complex spatial distribution of the electrophysiology properties of the heart, the whole cardiac geometry is decomposed into a set of coupled conductive media of different topology, namely (i) a 1D network of bundles comprising a fast conduction atrial network, the AV--node and the ventricular bundles; (ii) a 2D Purkinje network; and (iii) the 3D atrial and ventricular myocardium. 
These overlapping subdomains are two--way electrically coupled and the advancing depolarization front can propagate from one media to another, as happens in physiological conditions. Specifically, in a healthy heart, the  fast conduction atrial network activates the 3D atrial myocardium and the AV--node which, in turn, activates the ventricular bundles transmitting the depolarization front to the Purkinje network which rapidly activates the 3D ventricular myocardium through the PMJs. Nevertheless, different activation patterns, also including backward activation from the 3D myocardium to the bundles and/or to the Purkinje, may occur in pathological conditions as observed in section~ \ref{eq:patho}.
Although the propagation of the depolarization front in all these conductive media is governed by the bidomain/monodomain equations, the heterogeneity of the cardiac electrophysiology properties  at the cellular scale corresponds  to different electrical conductivities and ionic currents across the myocytes membrane at the continuum scale, which has been accounted in the numerical model through non--uniform conductivity tensors which depend on the local fiber orientation and using three different cellular models.
Specifically, the Courtemanche cellular model \cite{courtemanche1998ionic} is used for the atrial myocytes (and the corresponding internodal pathways), the Stewart cellular model \cite{stewart2009mathematical}  is adopted for the Purkinje Network, whereas the transmembrane ionic fluxes in the ventricular myocytes are solved through the ten~Tusscher--Panfilov cellular model \cite{ten2006}, which correctly reproduces the action potential within ventricular myocytes.
These models are coupled with the bidomain/monodomain equations, which are discretized in space using an in--house finite--volume method tailored  for 1D, 2D and 3D complex geometries and the explicit Rush--Larsen temporal integration scheme guarantees enhanced stability properties. The numerical solver has been thoroughly validated with available benchmark results from the literature \cite{niederer2011verification,cuccuru2015simulating} and the resulting depolarization within the whole heart well agrees with in--vivo observations  \cite{hall2015guyton}.
\\ \indent
\textcolor{black}{
The whole solver is GPU--accelerated using CUDA Fortran with the extensive use of kernel loop directives (CUF kernels)  providing an unprecedented speedup, thus allowing to solve a complete heartbeat in less than 8 wall--clock hours using Tesla V100 GPU devices.  
It should be noted that such computational cost could be further reduced either resorting to a monodomain model for the 3D myocardium corresponding to a 1.5 wall--clock hours per heartbeat, or using the next generation Tesla A100 GPU devices which are expected to provide a further acceleration of about four times while keeping the same code \cite{viola2021fsei}. }
\begin{figure}[!h]
\centering
\includegraphics[width=1\textwidth]{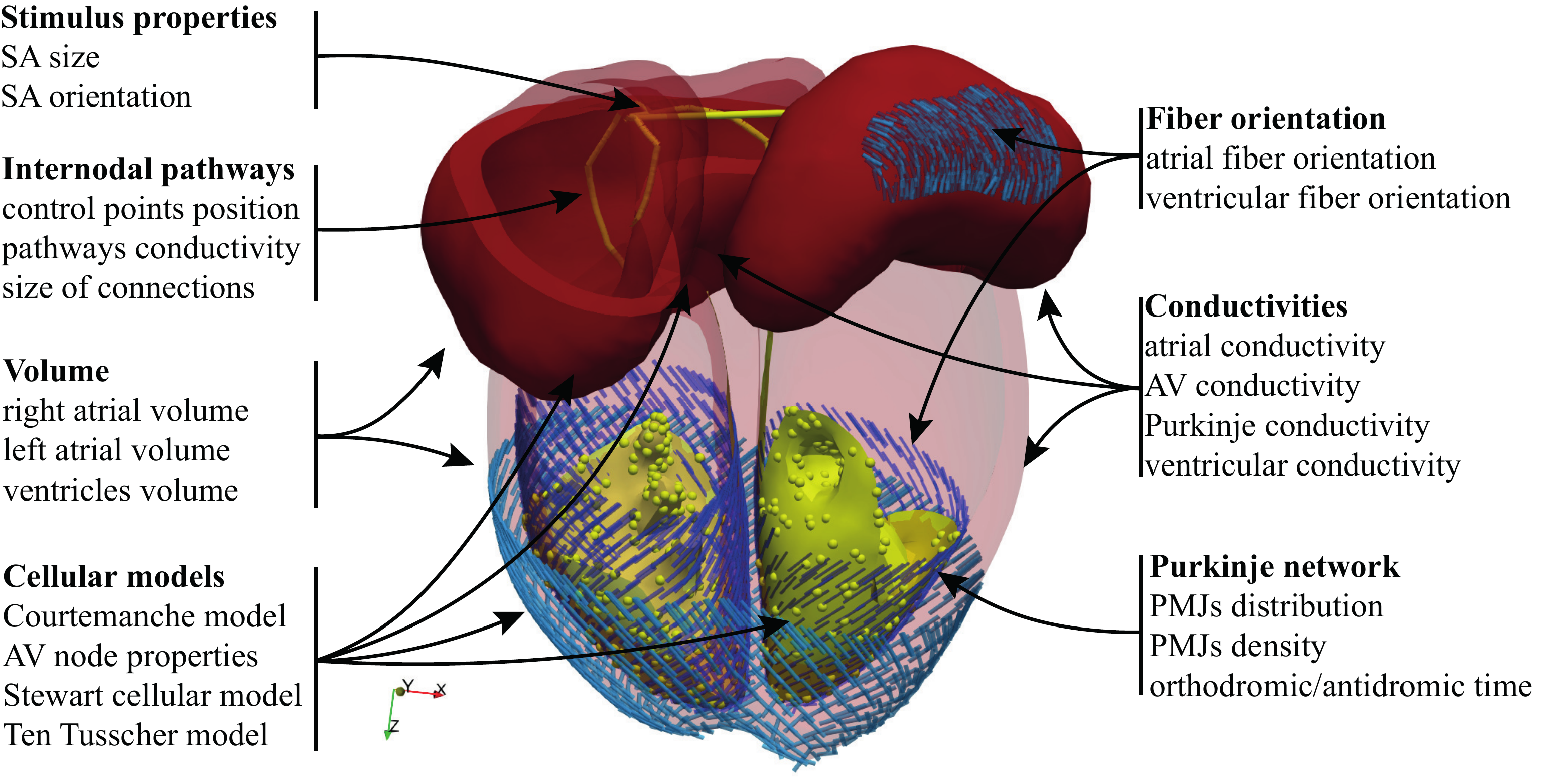}
\caption{Summary of the electrophysiology components having a high variability among individuals and which may be studied systematically with the proposed numerical model through an UQ analysis.}
\label{fig:UQfigure}
\end{figure}
\\ \indent Importantly, the computational high--performance of the solver and its versatility in controlling the geometrical and electrical properties of the whole cardiac electrophysiology open the way for systematic uncertainty quantification (UQ) analyses. The human electrophysiology system, indeed, presents high variability in the several of its components such as the fiber orientations, conductivity tensors, chambers volume, internodal pathways positions and the density of PMJs, and our numerical framework is designed to easily control and vary all these relevant quantities summarized in Figure~\ref{fig:UQfigure}.
The splitting of the cardiac electrophysiology system in a set of interconnected conductive media not only reduces the computational cost, but also provides an ideal framework for investigating the effect of an electrical or geometrical modification of the fast conduction network (bundles and/or Purkinje) on the cardiac depolarization, thus allowing to optimize cardiac resynchronization therapies or invasive surgical procedures \cite{lee2014applications, sun2014computational,lee2018computational}.
Furthermore, the computational bottleneck given by the 3D bidomain simulations can be circumvented by exploiting appropriate multi--fidelity strategies \cite{ng2012multifidelity,mollero2018multifidelity,fleeter2020multilevel}. 
On the other hand, a monodomain inverse conductivity problem (MICP) \cite{barone2020experimental,barone2020efficient} can be solved for the fast conduction network of bundles to calibrate the electrical conductivities of monodomain model in order to match medical data acquired in--vivo.
As a last  comment, the relationship between the cardiac valves functioning \cite{meschini2018effects,meschini2020heart,karas1970mechanism}
and the geometry of the Purkinje network, papillary muscles and chordae tendinae could be investigated by integrating our electrophysiology  model within a fluid--structure solver \cite{viola2020fluid}, so to also account for the cardiac hemodynamics and tissues kinematics in the UQ analysis.

\bibliography{mybibfile}

\end{document}